\documentclass[preprint,12pt]{elsarticle}

\usepackage{lmodern}

\usepackage[version=4]{mhchem}

\usepackage[english]{babel}
\usepackage[utf8]{inputenc}
\usepackage[T1]{fontenc}

\usepackage{sidecap}

\usepackage[makeroom]{cancel}

\usepackage[numbers]{natbib}
\biboptions{sort&compress}

\usepackage{graphicx}
\usepackage{epstopdf}
\usepackage{rotating}
\usepackage{wrapfig}
\usepackage{float}
\usepackage{subfigure}
\usepackage{color}

\usepackage{hyperref}
\usepackage{datetime}
\usepackage{amsmath,amssymb,amsthm}
\usepackage{amsfonts}
\usepackage{mathtools}
\usepackage{cases}
\usepackage{array}
\usepackage[binary-units=true]{siunitx}
\usepackage{upgreek}

\usepackage[title]{appendix}

\usepackage[stable]{footmisc}

\usepackage{booktabs}
\usepackage{multirow}
\usepackage{multicol}
\usepackage{caption}

\usepackage{bm}

\usepackage{natbib}
\usepackage{dirtree}
\usepackage[dvipsnames]{xcolor}
\usepackage{framed}
\usepackage{color}
\usepackage{emptypage}
\usepackage{verbatim} 
\usepackage{textcomp}
\usepackage[section]{placeins}
\usepackage{pdfpages}
\usepackage{enumitem}
\usepackage[flushleft]{threeparttable}
\usepackage{array,booktabs,makecell}
\usepackage{tcolorbox}
\usepackage{xargs}

\usepackage{marginnote}
\usepackage{lineno}

\usepackage[top=3.5cm, bottom=3.5cm, outer=3.5cm, inner=3.5cm, heightrounded, marginparwidth=3.cm, marginparsep=0.5cm]{geometry}

\setlist[enumerate]{itemsep=0mm}

\journal{Measurement}

\newcommand{\subrm}[1]{{_{\mathrm{#1} } }} 
\newcommand{\subsuprm}[2]{{ _{\mathrm{#1}}^{\mathrm{#2} } }} 

\newcommand{\planck}{\textrm{h}}
\newcommand{\speedoflight}{\textrm{c}}
\newcommand{\boltzmannconstant}{k_{\mathrm{B}}}
\DeclareSIUnit[per-mode=symbol,per-symbol=p]{\mbar}{\milli\bar}

\newcommand{\mdot}{\dot{m}_{\mathrm{gas}}} 
\newcommand{\Pel}{P_{\mathrm{el}}}
\newcommand{\Pelsim}{P_{\mathrm{el}}^{\textrm{sim}}}
\newcommand{\pc}{p_{\mathrm{c}}} 
\newcommand{\Twall}{T_{\mathrm{w}}} 

\newcommand{\IRMband}[2]{#1\textrm{-}\SI{#2}{\um}}
\newcommand{\spectral}{_{\lambda}}
\newcommand{\responsivity}[1][]{R_{\lambda_{#1}}}
\newcommand{\Rnorm}[1][]{\tilde{R}_{\lambda #1}}
\newcommand{\spectralradiance}[1][]{L_{\lambda}^{\textrm{#1}} }
\newcommand{\BBspectralradiance}{L_{\lambda}^{\textrm{bb}}}
\newcommand{\inband}[1][]{_{\Delta\lambda_{#1}}}
\newcommand{\throughput}{\Theta}
\newcommand{\tauwin}{ \tau_{\lambda}^{\textrm{win}}}
\newcommand{\tauatm}{ \tau_{\lambda}^{\textrm{atm}}}
\newcommand{\spectralemittance}[1][]{\varepsilon_{\lambda}^{\textrm{#1}}}
\newcommand{\spectraleps}{{\varepsilon_{\lambda}}}
\newcommand{\Tobj}{T_{\textrm{obj}}}
\newcommand{\Tapp}{T_{\textrm{app}}}

\newcommand{\Hfunction}{\mathcal{H}}

\newcommand{\cficp}[0]{{\texttt{CF-ICP}}}

\begin{document}

\begin{frontmatter}


\title{Line-of-sight gas radiation effects on near-infrared two-color ratio pyrometry measurements during plasma wind tunnel experiments}

\author[inst1,inst2]{Andrea Fagnani \corref{cor1}}
\ead{andrea.fagnani@vki.ac.be}
\cortext[cor1]{Corresponding author}
\author[inst1]{Bernd Helber}
\ead{bernd.helber@vki.ac.be}
\author[inst2]{Annick Hubin}
\ead{annick.hubin@vub.be}
\author[inst1]{Olivier Chazot}
\ead{olivier.chazot@vki.ac.be}

\affiliation[inst1]{organization={Aeronautics and Aerospace Department, von Karman Institute for Fluid Dynamics},
            addressline={Chaussée de Waterloo, 72}, 
            city={Rhode-st-Genèse},
            postcode={1640}, 
            country={Belgium}}
\affiliation[inst2]{organization={Materials and Chemistry Department, Vrije Universiteit Brussel},
       	addressline={Pleinlaan, 2}, 
       	city={Brussel},
       	postcode={1150},
       	country={Belgium}}

\begin{abstract}
Two-color ratio pyrometry is commonly used to measure the surface temperature of aerospace materials during plasma wind tunnel experiments.  However, the effect of the plasma radiation on the measurement accuracy is often neglected.  In this paper we formulate a model of the instrument response to analyze the systematic error induced by the gas radiation along the optical path. CFD simulations of the plasma flow field, together with a radiation code, allow to compute the gas spectral radiance within the instrument wavelength range. The measurement error is numerically assessed as a function of the true object temperature and emittance value. Our simulations explain the typical behavior observed in experiments, showing that a significant bias can affect the measured temperature during the material heating phase. For an actual experiment on a ceramic-matrix composite, a correction to the measured data is proposed, while comparative measurements with a spectrometer corroborate the results. 

\end{abstract}


\end{frontmatter}


\section{Introduction}
\label{sec:TCP_introduction}

The hypersonic flight of a spacecraft through a planetary atmosphere is a fascinating engineering endeavor, often representing the most critical part of a space mission. During atmospheric entry, the vehicle’s kinetic energy is converted into thermal energy of the flow across a strong shock wave, creating an extreme aero-thermal environment \cite{Anderson2006}.  Gas temperatures can rise above 10000~K, thus producing a chemically-reacting plasma flow.
Thermal Protection Systems (TPSs) are designed to shield the spacecraft from the severe heat loads \cite{Duffa2013}, ensuring safety of the crew and payload onboard.
In this context, Plasma Wind Tunnels (PWTs) offer a laboratory testing environment that allows to study complex gas-surface interaction phenomena, resulting from the TPS material response to high-temperature chemically-reacting flows \cite{Chazot2015}. These facilities typically use arc heaters or Inductively Coupled Plasma (ICP) torches to produce a high-temperature plasma jet, which is impinged onto a test material sample to duplicate a real flight scenario \cite{Turchi2021}.

In the context of PWT experiments, InfraRed (IR) radiometry is commonly used to characterize the material response during exposure to the plasma jet \cite{GulhanA1999}, as it allows non-intrusive probing of the surface temperature up to extreme values ($ \sim \SI{3000}{\kelvin} $). 
In particular, multi-spectral radiometry is a class of IR techniques that offers some key advantages. Single band IR measurements, in fact, require prior knowledge of the material's emittance to correct the measured signal in order to retrieve the target temperature \cite{Chrzanowski1995}. Multi-spectral systems, instead, provide simultaneous measurements of the object irradiance in $ N $ wavelength bands. By making suitable assumptions on the behavior of the spectral emittance, $ \varepsilon\spectral $, the system of measurement equations can be solved for the object temperature and the $ N-1 $ parameters describing $ \varepsilon\spectral $ \cite{Araujo2017, Huang2023, Zhang2022}. As a result, these techniques allow measuring the material's surface temperature even when its emittance is unknown or changing in an unpredictable fashion, which represents a key factor for PWT metrology \cite{DeCesare2020}.

Two-Color Pyrometry (TCP) is the simplest case of multi-spectral radiometry, employing only two measurements bands, and it is therefore widely employed to monitor the evolution of the material surface temperature during PWT experiments \cite{Panerai2012a, Helber2016b, Fagnani2023a, Milos2018, DeCesare2020, GulhanA1999, Bohrk2012, Luo2016}. Recently, some authors have proposed advanced applications, including two-color imaging pyrometry using CCD cameras \cite{Zander2016, Monier2017}, two-color thermographic imaging using IR cameras \cite{Musto2016, Savino2020, DiCarolo2020} and high-speed three-dimensional tomographic two-color pyrometry of flames \cite{Yu2021}.
Although one parameter can be used to model $ \varepsilon\spectral $ in TCP, it is common practice to assume a uniform behavior (gray-body) in the wavelength range of interest. As a result, the ratio of the measured signals becomes independent of the emittance, allowing to solve directly for the object temperature (two-color ratio pyrometry) \cite{Araujo2017}. In this case, the sensitive wavelength bands are typically selected very close to each other to reduce the approximation error (local gray-body assumption). This, however, increases the sensitivity of the signal ratio to noise and effects of participating media along the optical path, as well as to any variation of $ \varepsilon\spectral $, should it occur \cite{Tenney1988}. 
Accurate selection of the wavelength bands, in terms of range and spacing, determines the sensitivity of the instrument to a change in temperature of the emitting source \cite{Rodiet2016}. For values between $ \SI{500}{\kelvin} $ and $ \SI{2500}{\kelvin} $, such as those typically encountered on materials during PWT experiments, the highest sensitivity is obtained around $ \SI{1}{\um} $. 
This range is also weakly affected by optical path absorption from $ \ce{CO2} $ and $ \ce{H2O} $ \cite{Huang2019}, making it ideal for calibration purposes. 

The effects on TCP related to the gray-body assumption, in case of a real surface with variable spectral emittance \cite{Tapetado2016,  Musto2016}, as well as to environmental factors, in terms of ambient reflections \cite{Ketui2016, Saunders2000}, window transmission \cite{Lowe2015} or absorption along the optical path \cite{Huang2019, Mlacnik2021}, as well as background radiation \cite{Araujo2016}, have received considerable attention in the literature. However, the influence of gas radiation in PWT metrology is often neglected.
In this case, the plasma jet, reaching temperatures within $ 5000\textrm{-}\SI{10000}{\kelvin} $, produces an intense emission along the optical path between the instrument and the material's surface. 
This primarily originates from the spontaneous emission of excited atoms and molecules in the ultraviolet (UV) to near-infrared (NIR) wavelength region of the electromagnetic spectrum \cite{Laux2003, Fagnani2020a}. As this overlaps with the wavelength range of many commercial two-color pyrometers, we can expect the plasma emission to induce systematic errors on the measured temperature.  

Notably, Loesener and co-workers~\cite{Loesener1993, Loesener1994} already considered this source of interference; the problem, however, was chiefly avoided either selecting a suitable spectral region where the gas radiation was weak enough, or measuring the temperature of the material on the back surface, in order to avoid the pyrometer optical path to cross the plasma jet.  MacDonalds~et~al.~\cite{MacDonald2015} also conducted a preliminary analysis of the error induced by the plasma radiation from an ICP torch on a single-band IR pyrometer. 
The problem was treated more extensively in the context of plasma-sprayed particle diagnostics, where some researchers analyzed the influence of both direct plasma emission and scattered light on the measurement of the particle temperature by means of TCP. For instance, Sakuta and Boulos~\cite{Sakuta1988} defined a thermal visibility factor, based on the ratio of the particle emission to the sum of particle and gas emission along the line of sight. A critical plasma temperature could be identified, depending on the particle surface temperature and emittance value, defining an acceptable measurement range. Gougeon and Moreau~\cite{Gougeon1993}  investigated the influence of the scattered light by combining spectroscopic measurements and the Mie scattering theory. They showed that large positive errors could affect the measured temperature, as later confirmed by Salhi~et~al.~\cite{Salhi2005}. Correspondingly, a minimum measurable temperature could be defined as a function of the plasma emission intensity. More recently, Aziz~et~al.~\cite{Aziz2017} measured the spectral signature of an $ \ce{Ar/He} $ plasma loaded with zirconia particles, also concluding that TCP measurements could be significantly affected by the plasma emission along the line of sight.

Starting from these considerations, we propose to study the effect of the plasma emission along the optical path on TCP measurements during PWT experiments of TPS materials. Our approach is based on a model for the response of a two-color ratio pyrometer (sec.~\ref{sec:TCP_IRM}), which, after calibration (sec.~\ref{sec:TCP_setup}), allows to compute the effect of the gas radiation on the measured temperature. Starting from CFD simulations of the flow field, the plasma emission along the line of sight is numerically simulated with a radiation code. The effect on the measured temperature is studied as a function of the ICP torch electric power, as well as in terms of the material surface temperature and emittance value (sec.~\ref{sec:TCP_optical_path}), showing that a large positive bias can be expected during the transient heating phase of the material sample. Experimental results on a ceramic matrix composite (sec.~\ref{sec:TCP_results}), tested in the Plasmatron facility at the von Karman Institute for Fluid Dynamics (VKI), confirm the predicted trends. A correction to the measured signals allows to compensate for the biasing effect, while a comparative analysis with a spectrometer, simultaneously probing the line-of-sight emission, corroborates the results.

\section{TCP Instrument Response Model}
\graphicspath{{figures/}}
\label{sec:TCP_IRM}

Let us consider an IR instrument with absolute spectral radiance responsivity
$ \responsivity~=~k\Rnorm[]~[\SI{}{\ampere \per \watt}] $, which includes detector sensitivity and internal optics transmission, where $ \Rnorm $ is the normalized spectral response and $ k $ is the absolute sensitivity coefficient. Then, the radiometric measurement equation relates the instrument output signal, $ S $, to the spectral radiance incident on the instrument collector optics, $ \spectralradiance $, as \cite{Wyatt1978, Nutter1988}
	\begin{equation}
		S =  \throughput k \int\inband \Rnorm \spectralradiance d\lambda \;\; [\SI{}{\ampere}],
		\label{eq:TCP_signal}
	\end{equation}
where $ \throughput \; [\SI{}{\meter\squared \steradian}] $ is the instrument optical throughput, $ \lambda $ is the wavelength and $ \Delta\lambda $ is the range defined by $ \Rnorm $. Then, the signal S is digitalized by the instrument electronics and recorded through a software interface.

\begin{figure}[]
	\centering
	\includegraphics[trim = {0cm, 6.5cm, 8cm, 0cm}, clip, width=.7\textwidth]{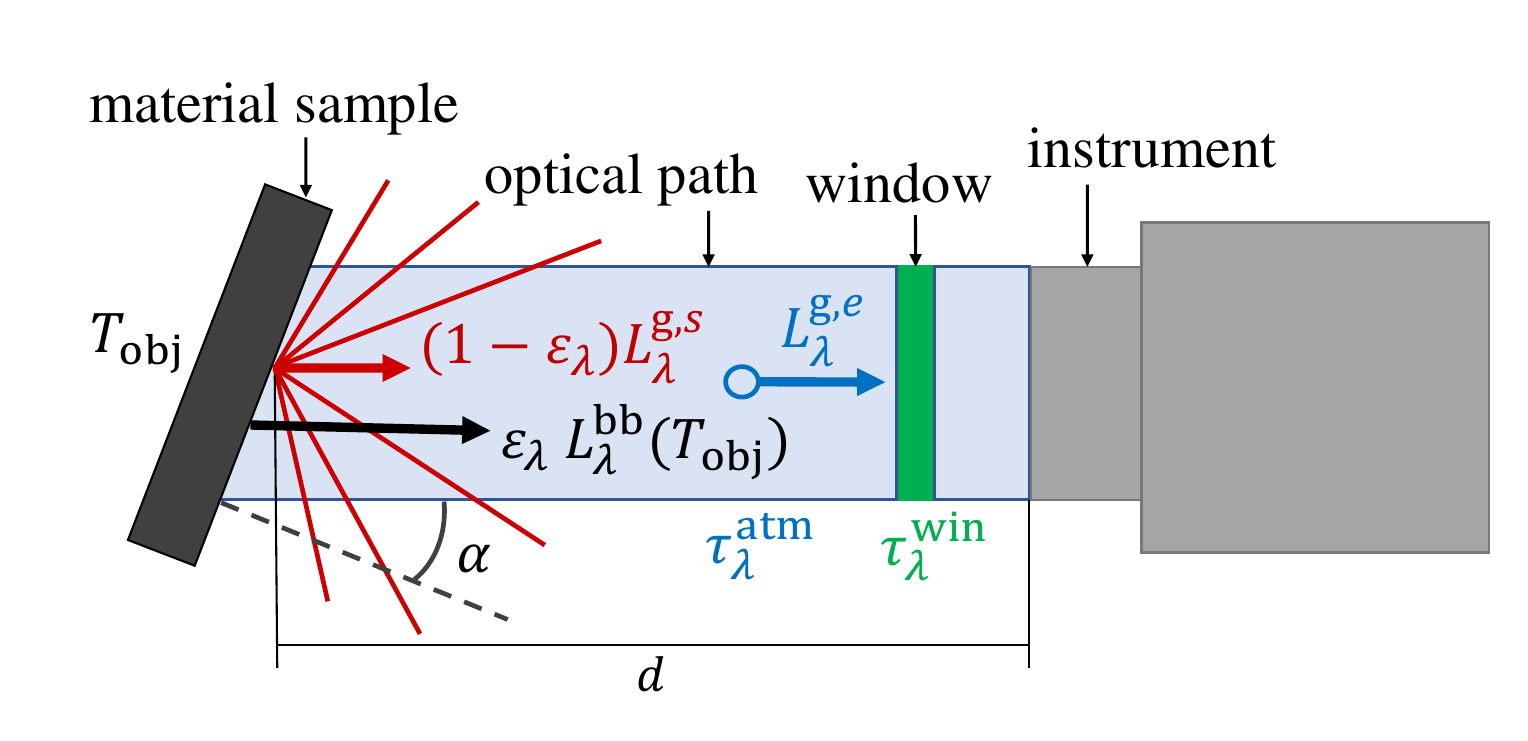}
	\caption{Schematic of the contributions to the radiance detected by the instrument. The collection volume is considered to be a cylinder of size $ A \times d $, incident on the material surface with an angle $ \alpha $ with respect to the surface normal.}
	\label{fig:sketch_radiance}
\end{figure}

Neglecting ambient reflections and optics emission for the high object temperatures of our application ($ \Tobj > \SI{500}{\kelvin} $), considering the schematic in Fig.~\ref{fig:sketch_radiance}, $ \spectralradiance $ can be written as
	\begin{equation}
		\begin{split}
				\spectralradiance &\approxeq \tauatm \tauwin \varepsilon\spectral \spectralradiance[bb] (\Tobj) + \tauwin \spectralradiance[g,e] + \\
				&+ \tauatm \tauwin (1-\spectraleps) \spectralradiance[g,s] \;\; [\SI{}{\watt \per \meter\squared \per \steradian \per \um}].
		\end{split}
		\label{eq:TCP_spectral_radiance}
	\end{equation}
Here, $ \tauatm $ and $ \tauwin $ represent the spectral transmittance of the atmosphere and external optics along the optical path, respectively, while $ \varepsilon\spectral $ is the spectral directional emittance of the material's surface.  $ \spectralradiance[bb] (\Tobj) $ is the spectral radiance of an ideal blackbody at temperature $ \Tobj $, described by Planck's law as \cite{Modest2013}
\begin{equation}
	\BBspectralradiance(\lambda, T) = \frac{2 \planck \speedoflight}{\lambda^5} \frac{1}{\exp[\planck \speedoflight / (\boltzmannconstant \lambda T)]-1},
\end{equation} 
where $ \planck $ is the Planck's constant, $ \speedoflight $ is the speed of light and $ \boltzmannconstant $ is the Boltzmann's constant. 
Lastly, $ \spectralradiance[g,e] $ is the spectral radiance emitted by the gas along the optical path, while $ (1-\varepsilon\spectral)\spectralradiance[g,s] $ is the spectral radiance emitted by the surrounding gas and reflected by the material's surface onto the instrument collector optics.   

In this work, we focus on the influence of $ \spectralradiance[g,e] $ on the measured temperature; hence, the following assumptions are introduced. 
Considering highly emitting materials, i.e.,  $ \spectraleps \geq 0.8 $, the contribution of $\spectralradiance[g,s] $ in the previous equation is considered negligible. 
However, we remark that this could become important in case of highly reflective surfaces.
Moreover, for an instrument with operating range around $ \SI{1}{\um} $ and a stand-off distance of $ \SI{1}{\meter} $, we consider a thin optical path ($ \tauatm \approxeq 1 $). Finally, we also assume that the window material is carefully chosen to provide a uniform spectral transmittance in the instrument wavelength range, such that $ \tauwin = \tau\subrm{win} $. 

Inserting eq.~\ref{eq:TCP_spectral_radiance} in eq.~\ref{eq:TCP_signal}, and considering the aforementioned assumptions, the ratio, $ \rho\subrm{m} $, of the signals, $ S_1 $ and $ S_2 $, output from each wavelength band of a two-color pyrometer writes
	\begin{equation}
		\rho\subrm{m} = \frac{S_1}{S_2} \approxeq \frac{k_1}{k_2} \times \frac{  \int\inband[1]  \Rnorm[,1] [\spectraleps \spectralradiance[bb] (\Tobj) + \spectralradiance[g,e] ] d\lambda}{ \int\inband[2] \Rnorm[,2] [\spectraleps \spectralradiance[bb] (\Tobj) + \spectralradiance[g,e]] d\lambda},
		\label{eq:TCP_rho_meas}
	\end{equation}
where the factor $ \varphi = k_1 / k_2 $ accounts for the absolute sensitivity ratio between $ \responsivity[,1] $ and $ \responsivity[,2] $. Since only the normalized responses $ \Rnorm[, i] $ are generally provided by the manufacturer, this coefficient can be determined during calibration, as later discussed in section~\ref{sec:TCP_calibration}. The previous equation also considers the same optical throughput for the two measuring bands, i.e., $ \throughput_1 = \throughput_2 $, which is the case for dual sandwich detectors considered in this work. 

During calibration, performed in a controlled laboratory environment at room temperature, optical path emission is negligible, i.e., $ \spectralradiance[g,e] \approx 0 $, and a blackbody calibration source provides $ \spectraleps \approxeq 1 $. Hence, eq.~\ref{eq:TCP_rho_meas} reduces to 
	\begin{equation}
		\rho\subrm{c} = \varphi \frac{ \int\inband[1] \Rnorm[,1]\spectralradiance[bb] (\Tobj) d\lambda }{ \int\inband[2] \Rnorm[,2] \spectralradiance[bb] (\Tobj) d\lambda } = \mathcal{H}\subrm{c}(\Tobj),
		\label{eq:TCP_Hfunction}
	\end{equation}
where $ \rho\subrm{c} $ indicates the signal ratio obtained during calibration and $ \Hfunction\subrm{c} $ represents the calibration curve.
During measurements in a PWT, instead, emission from the high-temperature plasma along the optical path will affect the measured ratio when $ \spectralradiance[g,e] \approx \varepsilon\spectral \spectralradiance[bb](\Tobj) $. In this case, as $ \rho\subrm{m} $ will differ from $ \rho\subrm{c} $ for the same value of $ \Tobj $, one only measures the apparent temperature
	\begin{equation}
		\Tapp = \Hfunction\subsuprm{c}{-1}(\rho\subrm{m}),
		\label{eq:TCP_Tapp}
	\end{equation}
and the quantity
	\begin{equation}
		e = \frac{\Tapp - \Tobj}{\Tobj} \times 100
		\label{eq:TCP_error}
	\end{equation}
will describe the relative measurement error percent in the following analysis. Eq.~\ref{eq:TCP_rho_meas}, eq.~\ref{eq:TCP_Hfunction} and eq.~\ref{eq:TCP_Tapp} represent the Instrument Response Model (IRM) that will be used to study the effect of the gas emission along the line of sight on the temperature measured by means of TCP.  
For this, we consider a target material with uniform spectral emittance in the instrument range, in order to satisfy the gray-body assumption of ratio pyrometry.

\section{Experimental set-up and calibration}
\graphicspath{{figures/}}
\label{sec:TCP_setup}

\subsection{The VKI Plasmatron facility}
\begin{figure}[]
	\centering
	\subfigure[]
	{\includegraphics[trim={7.5cm, 15cm, 18cm, 1.6cm}, clip, width=0.42\textwidth]{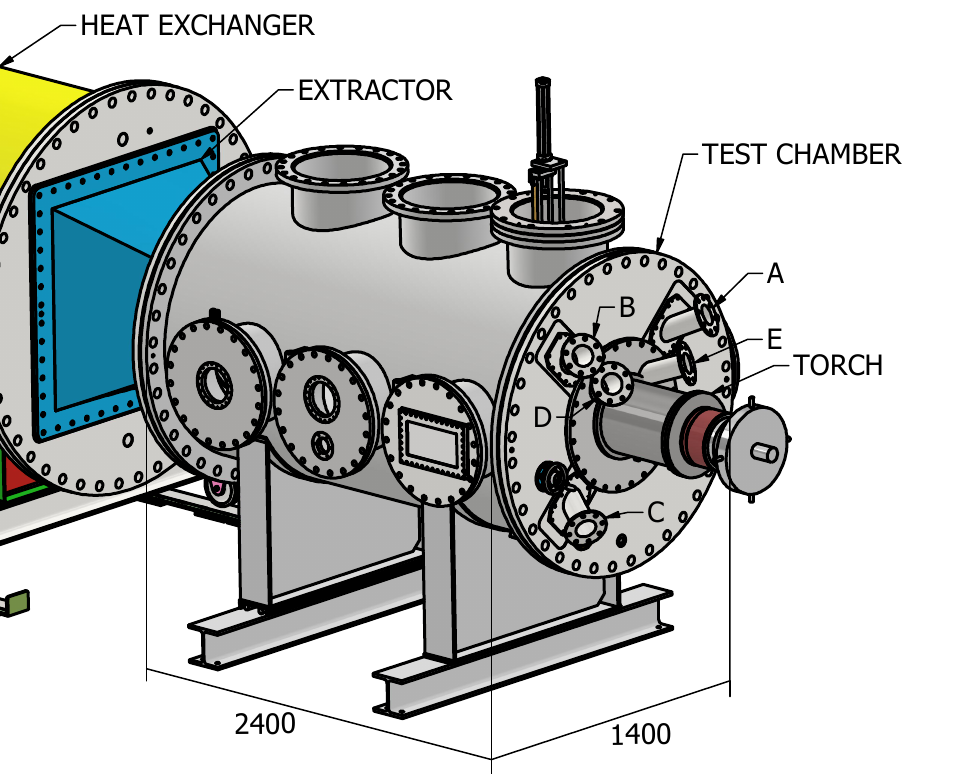} \label{fig:TCP_plasmatron_rendering}}
	\subfigure[]
	{\includegraphics[trim={2.5cm, 9cm, 12cm, 1.5cm}, clip, width=0.55\textwidth]{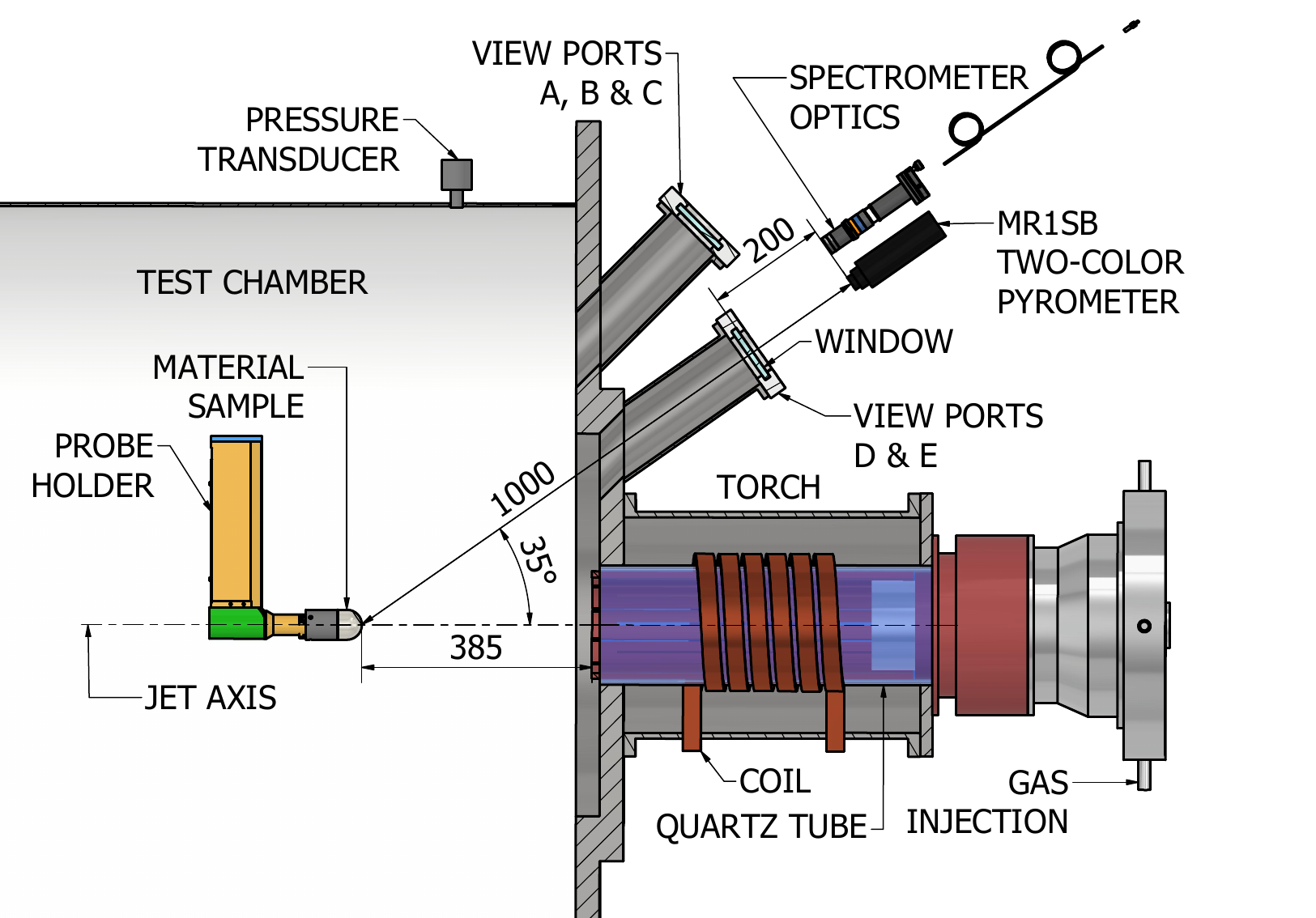} \label{fig:TCP_sketch_section}}
	\caption[]{(a) Rendering of the VKI Plasmatron facility, showing the test-chamber and the location of the view ports. (b) Schematic of the experimental set-up, highlighting the main components of the ICP torch and the geometry of the optical paths of the two-color pyrometer and spectrometer. Dimensions in millimeters.}
\end{figure}

The VKI Plasmatron facility, rendered in Figure~\ref{fig:TCP_plasmatron_rendering}, is equipped with a $ \SI{160}{\milli\meter} $ diameter ICP torch, powered by a $ \SI{400}{\kilo \hertz} $, $ \SI{1.2}{\mega \watt} $, $ \SI{2}{\kilo \volt} $ electric generator and connected to a $ \SI{1.4}{\meter} $ diameter, $ \SI{2.4}{\meter} $ long test chamber. An extensive description of the facility and its performance was given by Bottin and co-workers \cite{Bottin1999, Bottin1997a, Bottin1999a}. Fig.~\ref{fig:TCP_sketch_section} shows a schematic section of the test chamber and ICP torch. The latter is made up of a quartz tube, surrounded by a six-turn flat coil inductor and supplied by a gas injection system. The electric power to the coil, $ \Pel $, is monitored by a voltage-current probe, while a calibrated flow meter (F-203AV, Bronkhorst High-Tech B.V, NL) controls the mass flow rate, $ \mdot $, of the test gas supplied to the torch. The gas, compressed atmospheric air in this case, is heated by electromagnetic induction, thus providing a chemically pure plasma flow. Pressure in the test chamber, $ \pc $, is measured by an absolute pressure transducer (Membranovac DM 12, Leybold GmbH, DE). The material sample is mounted onto a movable probe holder, which can be injected into the plasma flow at a distance of $ \SI{385}{\mm} $ from the torch exit. 

\subsection{Two-color pyrometer and calibration}
\label{sec:TCP_calibration}

\begin{figure*}[]
	\centering
	\subfigure[]
	{\includegraphics[trim={0cm, 0cm, 0cm, 0cm}, clip, width=0.38\textwidth]{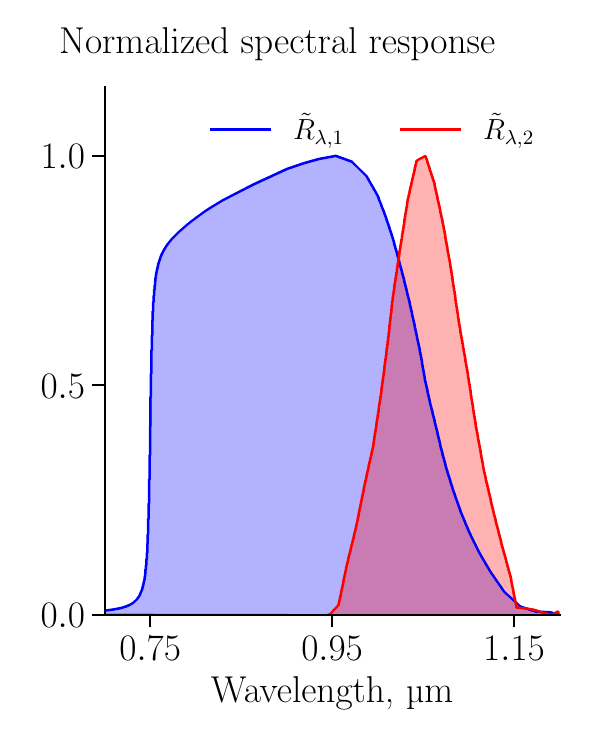} \label{fig:response}}
	\subfigure[]
	{\includegraphics[trim={0cm, 0cm, 0cm, 0cm}, clip, width=0.58\textwidth]{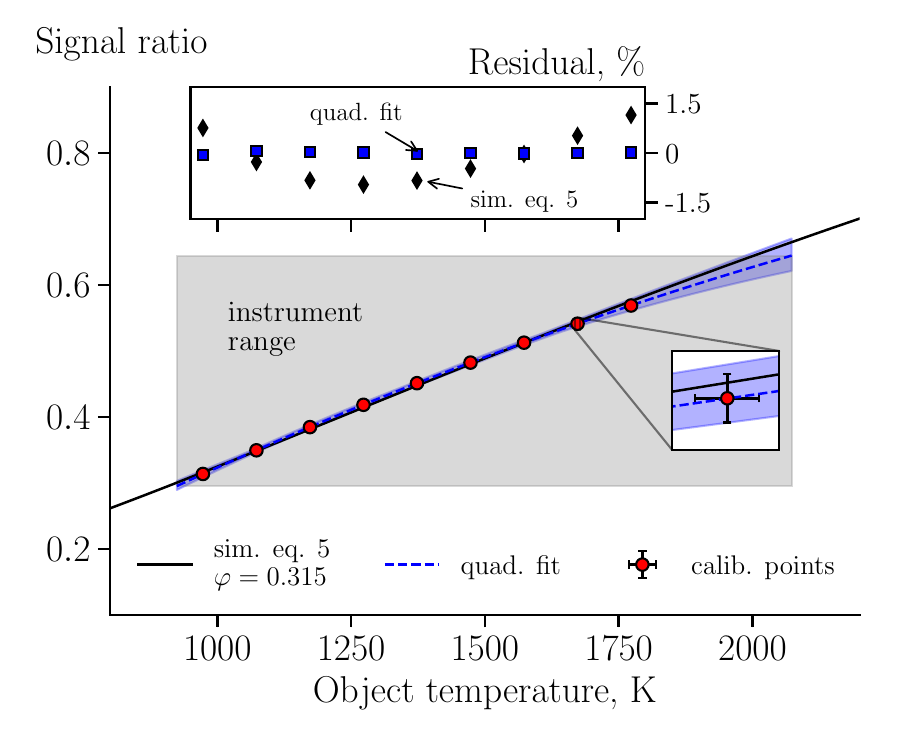} \label{fig:MR1SB_AF2021A_IRCC2CTP}}
	\caption[]{(a) Normalized spectral responses, $ \Rnorm[, 1] $ and $ \Rnorm[, 2] $, of the Raytek Marathon Series MR1SB two-color pyrometer.  (b) Simulated signal ratio from eq.~\ref{eq:TCP_Hfunction} closely matches the calibration points to a maximum residual of $ 1.5\% $, after fitting the modulation factor ($ \varphi~=~0.315 $). A quadratic function fits the calibration points with negligible residuals and will be used to process the actual experimental data in sec.~\ref{sec:TCP_results}.}
\end{figure*}

We used the Marathon Series MR1SB (Raytek Corporation, USA) two-color pyrometer, featuring a sandwich silicon detector with spectral responsivities between $ 0.75\textit{-}\SI{1.15}{\micro\meter} $ and $ 0.95\textit{-}\SI{1.15}{\micro\meter} $ (Fig.~\ref{fig:response}) and a temperature range between $ \SI{973}{\kelvin} $ and $ \SI{2073}{\kelvin} $. Signals were recorded with an integration time of $ \SI{10}{\ms} $ and at a frequency of $ \SI{10}{\hertz} $. Optical access to the Plasmatron test chamber was provided through a $ \SI{1.5}{\centi \meter} $ thick quartz window (label D in Fig.~\ref{fig:TCP_plasmatron_rendering}), whose spectral transmission in the instrument range can be considered uniform with a value $ \tau_{\textrm{win}} \approxeq 0.87 $. The instrument is placed at $ \SI{1}{\meter} $ distance from the sample and the optical access provides an inclination of about $ \SI{35}{\degree} $ with respect to the surface normal. The probing area over the sample surface has a size of approximately $ \SI{14}{\milli\meter} $ in diameter.  

The radiometric calibration of the instrument was performed with a variable temperature blackbody source (R1500T, Ametek Landcal, UK) in the range $ 973\textit{-}\SI{1773}{\kelvin} $. The latter features a $ \SI{120}{\degree} $ conical-ended silicon-carbide cavity with a $ \SI{40}{\mm} $ clear aperture and a PID controller to hold the set temperature. The uncertainty on the source temperature is $ \pm \SI{3}{\kelvin} $, while its emittance is considered unitary and spectrally uniform in the instrument wavelength range ($ \IRMband{0.75}{1.15} $).

During calibration, the two-color pyrometer is placed in front of the calibration source, replicating the operating distance and position of the window along the optical path. Then, the signal output from each band is recorded for a set of source temperatures to provide the calibration points. Figure~\ref{fig:MR1SB_AF2021A_IRCC2CTP} shows the signal ratio, $ \rho_{\textrm{c}} $, as a function of $ \Tobj $. 
A value of $ \varphi = 0.315 $ allowed to best fit the simulated response through eq.~\ref{eq:TCP_Hfunction} to the calibration points, with a residual within $ \pm 1.5\% $. This deviation, although very limited, can be related to an imperfect knowledge of the sensor responsivities. 
While the simulated response will be used in sec.~\ref{sec:TCP_optical_path} to study the effect of $ \spectralradiance[g,e] $ on the measured temperature, 
a quadratic function fits the calibration points with negligible residuals and will be used in sec.~\ref{sec:TCP_results} to process the actual experimental data. The quadratic fit is extrapolated to $ \SI{2073}{\kelvin} $ to cover the whole instrument range.
Uncertainties are propagated through the processing steps with a Monte Carlo sampling approach. We consider $ \pm\SI{3}{\kelvin} $ on the calibration source temperature and $ \pm1\% $ on both $ S_1 $ and $ S_2 $, including resolution and repeatability. This leads to about $ \pm1.5\% $ uncertainty on the measured temperature, excluding possible biasing effects.

\subsection{Spectrometer and calibration}
\label{sec:TCP_setup_spectro}
\begin{figure*}[h]
	\centering
	\subfigure[]
	{\includegraphics[trim = {0cm, 3cm, 1cm, 0cm}, clip, width=.65\textwidth]{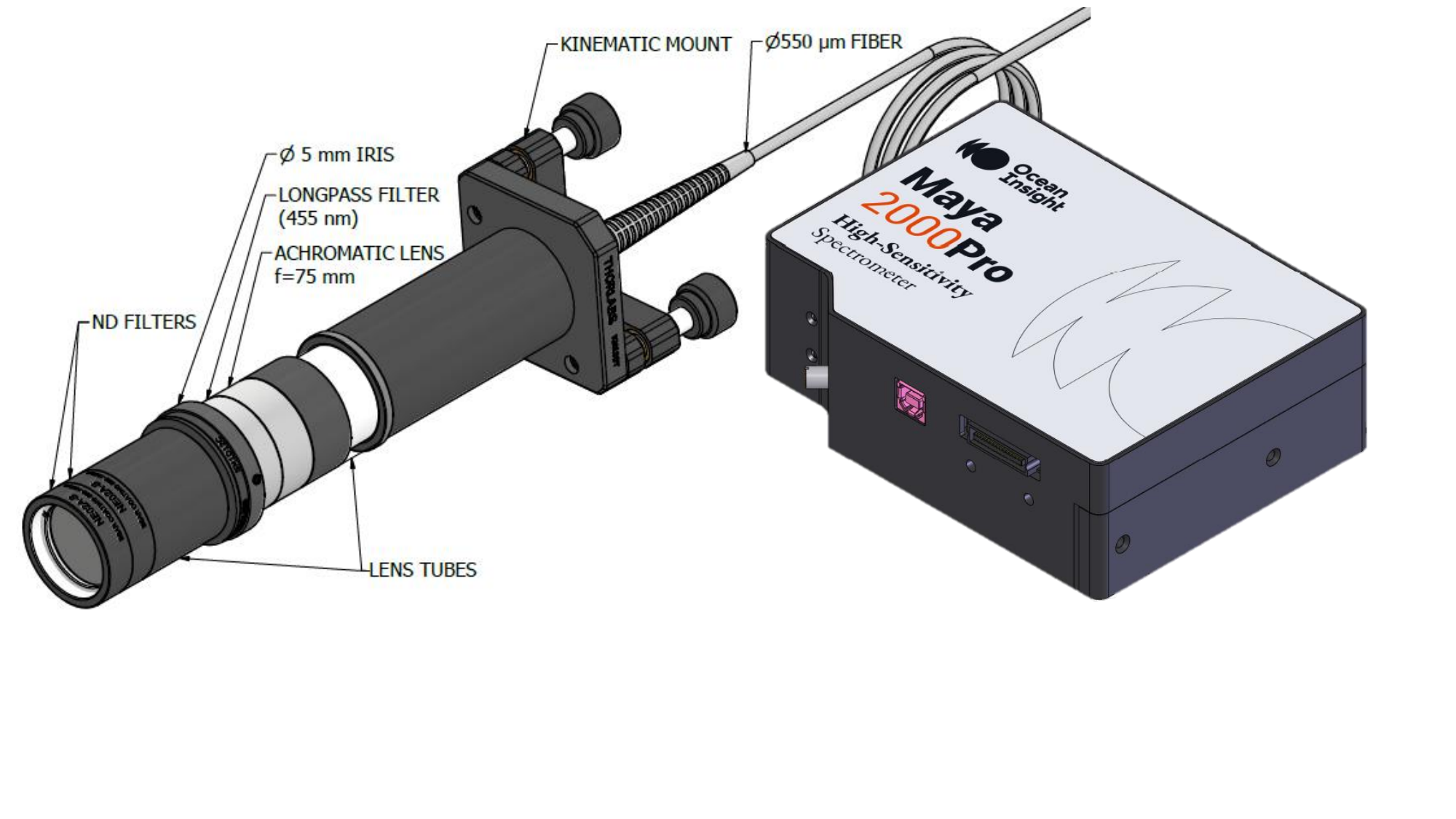} \label{fig:SURFspectro}}
	\subfigure[]
	{\includegraphics[trim={0cm, 0cm, 0cm, 0cm}, clip, width=0.30\textwidth]{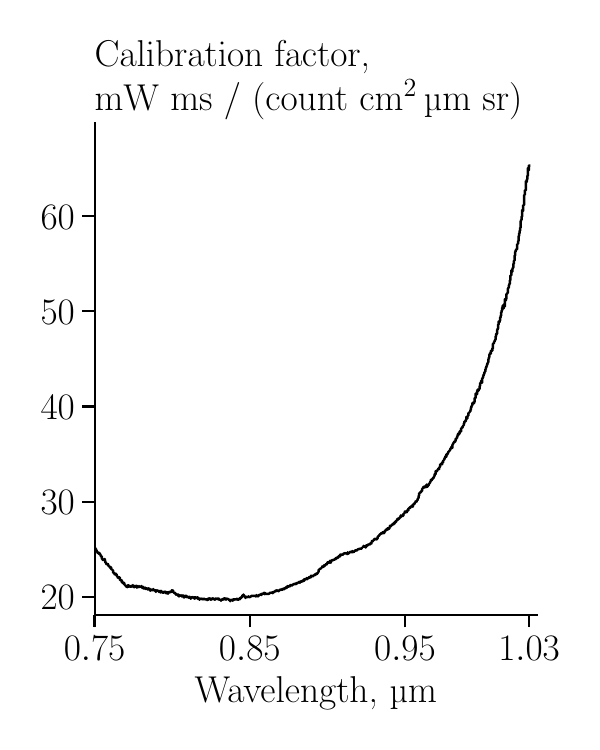} \label{fig:SURFspectro_AF2023A_calib_factor}}
	\caption[]{(a) Rendering of the compact spectrometer, together with the optical set-up designed for this work. (b) Calibration factor of the spectrometer as a function of wavelength.}
\end{figure*}

We used a compact spectrometer (MAYA 2000 PRO series, Ocean Insight, USA), coupled to a custom collection optics, as shown in Fig.~\ref{fig:SURFspectro}, to probe the spectral radiance emitted by the surface of the material sample within the $ 750-\SI{1030}{\nano \meter} $ range. 
The spectrometer features a $ \SI{25}{\um} $ wide entrance slit, a $ \SI{300}{grooves/\mm} $ grating and a back-thinned 2D CCD detector. 
The integration time was set to $ \SI{1}{\second} $ and the instrument was triggered at a frequency of $ \SI{0.5}{\hertz} $ by a pulse generator (DG535, Stanford Research Systems, USA), allowing precise synchronization with the two-color pyrometer.

The collection optics consists of a $ \SI{550}{\um} $ core diameter multi-mode optics fiber, a $ \SI{75}{\mm} $ focal length achromatic lens, a $ \SI{455}{\nm} $ high-pass filter and a $ \SI{5}{\mm} $ diameter iris. 
The optics are mounted on a $ \SI{1}{inch} $ diameter tube, held by a kinematic mount for precise alignment to the material surface. A stack of Neutral Density (ND) filters is additionally mounted to decrease the detected irradiance and avoid saturation of the detector. 
Optical access to the Plasmatron chamber was provided through a $ \SI{5}{\mm} $ thick $ \ce{CaF2} $ window (label E in Fig.~\ref{fig:TCP_plasmatron_rendering}), offering $ \sim 95\% $ transmission in the wavelength range. In a similar configuration to the two-color pyrometer, the collection optics are placed at $ \sim \SI{100}{\centi \meter} $ from the sample surface with a $ \sim 35^{\circ} $ inclination with respect to its normal. The probing area has a size of about $ \SI{10}{\mm} $ in diameter on the sample surface. The spectral resolution was characterized with a low-pressure $ \ce{Ar} $ lamp, resulting in about $ \SI{0.75}{\um} $ full-width at half maximum.

The system was calibrated using the same reference blackbody source described previously, for a source temperature of 1773~K and reproducing the optical path encountered in the measurement. 
Indicating with $ \hat{U}_{\lambda} = U_{\lambda} - U_{\lambda, \textrm{bg}} $ the raw signal detected by the spectrometer, $ U_{\lambda} $, minus the background signal, $ U_{\lambda, \textrm{bg}} $,  (in digital intensity counts) and with $ \Delta t $ the integration time, the measured spectral radiance $ \spectralradiance[m] $ is obtained as \cite{Kunze2009}
	\begin{equation}
		\spectralradiance[m] = \frac{ \hat{U}\subsuprm{\lambda}{m} } {\Delta t\subrm{m}} \cdot f\subsuprm{\lambda}{c},
		\label{eq:spectro_calib}
	\end{equation}
where
	\begin{equation}
		f\subsuprm{\lambda}{c} = \frac{\spectralradiance[bb]} { \hat{U}\subsuprm{\lambda}{c} / \Delta t\subrm{c} } \;\; [\SI{}{\milli \watt \; \milli \second / (count \; \centi\meter\squared \; \um \; \steradian )}]
	\end{equation}
is the calibration factor and the letters "m" and "c" indicate the quantities recorded during measurement and calibration, respectively. Figure~\ref{fig:SURFspectro_AF2023A_calib_factor} shows the calibration factor as a function of wavelength. 
The calibration law assumes linearity of the signal $ \hat{U}_{\lambda} $ with respect either to the incident radiance and integration time, which was checked during calibration. 

\section{Numerical assessment of the measurement error due to plasma emission along the line of sight}
\graphicspath{{figures/}}
\label{sec:TCP_optical_path}

\subsection{Plasma flow field}

The subsonic plasma flow in the Plasmatron chamber was numerically simulated using a two-dimensional magnetohydrodynamic solver, referred to as $ \cficp $ in the following, which couples the Maxwell induction equations with the Navier-Stokes equations under the assumptions of Local Thermodynamic Equilibrium (LTE) and axisymmetric steady flow \cite{Degrez2004}. The code is integrated into the Computational Object-Oriented Library for Fluid Dynamics (COOLFluiD) \cite{Lani2013} and relies on the Mutation++ library \cite{Scoggins2020} to compute the thermodynamic and transport properties of an eleven-species air mixture ($ \ce{O2} $, $ \ce{N2} $, $ \ce{O2+} $, $ \ce{N2+} $, $ \ce{NO} $, $ \ce{NO+} $, $ \ce{O} $, $ \ce{O+} $, $ \ce{N} $, $ \ce{N+} $, $ \ce{e-} $). Under well-established assumptions, the flow in the ICP torch can be considered continuum, partially ionized, and collision-dominated \cite{Magin2004a}. Then, the Navier-Stokes equations are used to express mass, momentum and energy conservation.  The electromagnetic field is modeled with a simplified form of Maxwell’s induction equation, coupled with the momentum and energy equations through Lorentz force and Joule heating effects. As the Reynolds number is typically low ($ Re \sim 100 $), the flow is assumed to be laminar and transition is neglected. The LTE model is adopted, where energy modes are assumed to follow a Maxwell-Boltzmann distribution and equilibrium chemistry occurs. 

\begin{figure*}[h!]
 	\centering
 	\subfigure[]
 	{\includegraphics[trim = {0.5cm, 1.8cm, 13.5cm, 0cm},clip, width=.50\textwidth]{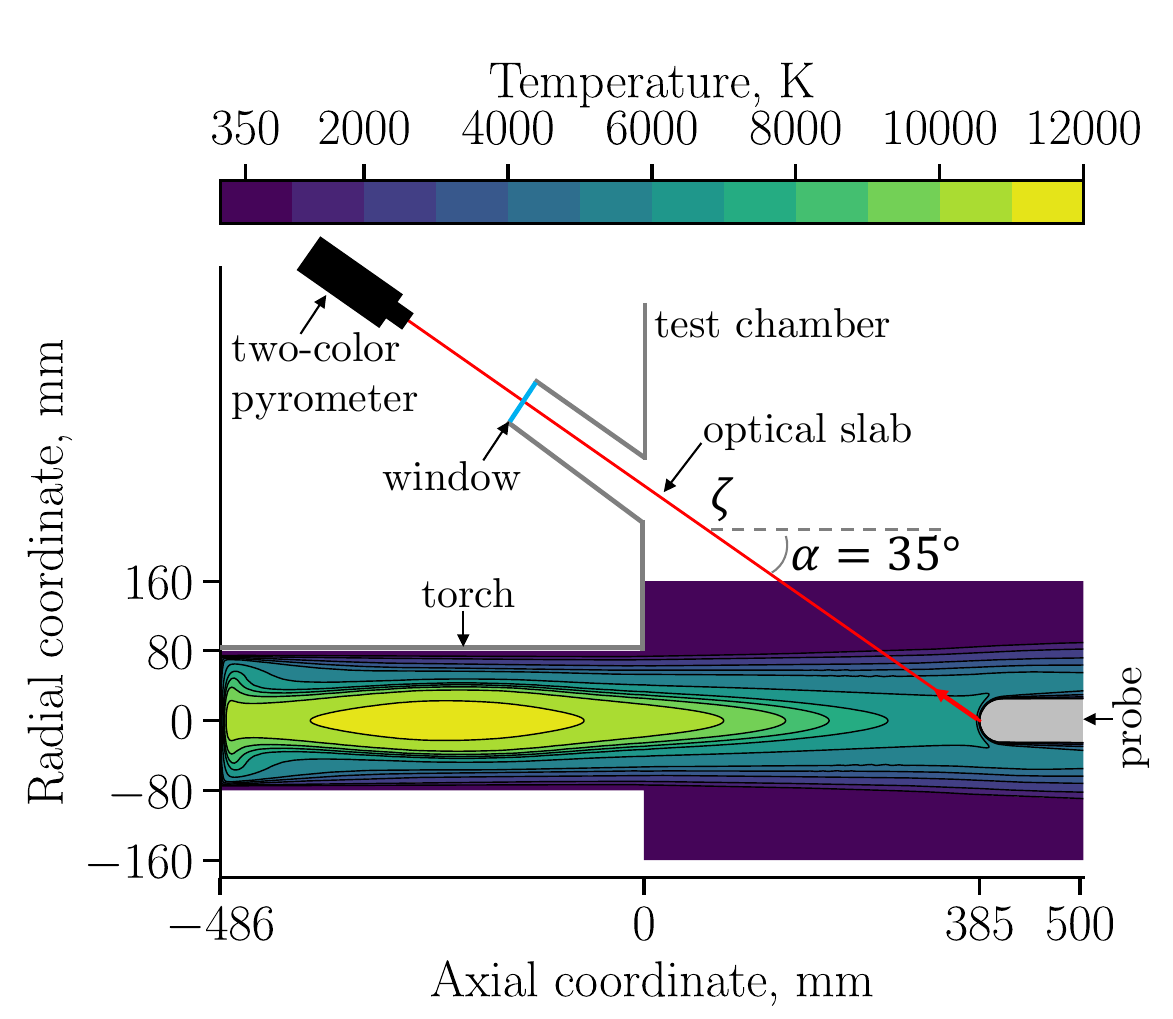}
 		\label{fig:ICP_PR_air_50mbar_90kW_HS50}}
 	\subfigure[]
 	{\includegraphics[trim = {0.5cm, 0cm, 0.5cm, 0cm}, clip, width=.45\textwidth]{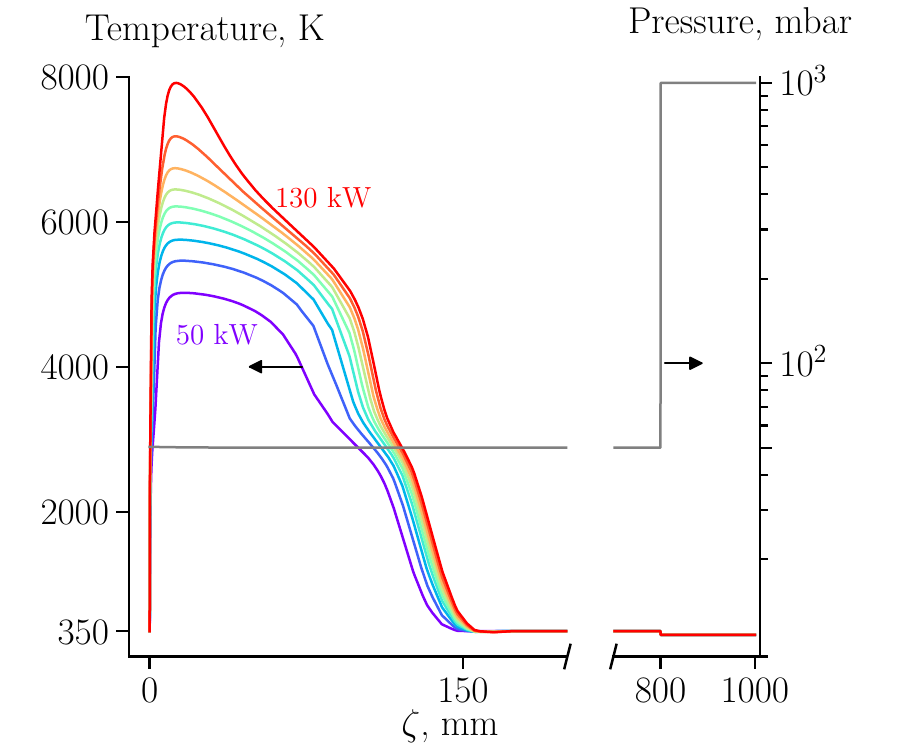} 	\label{fig:ICP_LoS_air11_50mbar_Pel50-130kW}}
 	\caption{(a) Temperature field of the VKI Plasmatron ICP torch, simulated with $ \cficp $ code for an eleven-species air mixture with $ \pc = \SI{50}{\mbar} $, $ \mdot = \SI{16}{\gram \per \second} $ and $ \Pelsim = \SI{90}{\kW} $. Also represented is the $ \SI{1}{\meter} $ optical slab from the material sample up to the instrument collection optics. (b) Temperature and pressure distributions along the optical slab, extracted from the CFD simulations as a function of $ \Pelsim $. Since the domain is limited to $ \SI{160}{\mm} $ from the jet axis, values are extrapolated up to the position of the window ($ \zeta = \SI{800}{\mm} $), while laboratory conditions ($ T = \SI{300}{\kelvin} $ and $ p = \SI{1}{atm} $) are considered for $ \SI{800}{\mm} < \zeta < \SI{1000}{\mm} $. }
 \end{figure*}

Figure~\ref{fig:ICP_PR_air_50mbar_90kW_HS50} shows an example of the simulated temperature field. The domain includes the torch and a portion of the test chamber up to $ \SI{500}{\mm} $ from the torch exit and $ \SI{160}{\mm} $ from the jet axis, as well as a representative $ \SI{50}{\mm} $ diameter hemispherical probe positioned at $ \SI{385}{\mm} $ downstream of the torch.
A quadrilateral mesh with $ \SI{16000}{cells} $ was used for the computation and a convergence study confirmed the grid independence. 
Boundary conditions are specified in terms of the torch and probe wall temperature ($\Twall = \SI{350}{\kelvin} $), chamber pressure ($ \pc = \SI{50}{\mbar} $),  inlet gas mass flow rate ($ \mdot = \SI{16}{\gram / \second} $) and input numerical electric power ($ \Pelsim $). 
Nine simulations were carried out for $ \SI{50}{\kilo \watt} < \Pelsim <  \SI{130}{\kilo \watt} $, with increasing steps of $ \SI{10}{\kilo\watt} $. 
In this regard, it is important to notice that $ \Pelsim $ differs from the value measured experimentally, $ \Pel $, due to the energy efficiency of the ICP torch. Recent comparison with spectroscopic temperature measurements of the gas at different axial locations suggested that  $ \Pelsim / \Pel \approxeq 35\textit{-}40\% $ \cite{Fagnani2020a}. 

The figure also represents a $ \SI{1}{\meter} $ long optical slab ($ \zeta\textrm{-axis}$), originating from the surface of the probe with a $ \SI{35}{\degree} $ inclination with respect to the jet axis and reaching the the two-color pyrometer. 
Figure~\ref{fig:ICP_LoS_air11_50mbar_Pel50-130kW} shows the gas temperature and pressure along the optical slab for the different values of $ \Pelsim $. 
Temperature increases rapidly through the boundary layer around the probe, up to several thousands of Kelvin, and drops to $ \SI{350}{\kelvin} $ outside the jet core, i.e., for $ \zeta > \SI{150}{\mm} $.  
Pressure, instead, is fairly uniform in the test chamber.

\newpage
Since the simulation domain terminates at $ \SI{160}{\mm} $ from the jet axis, corresponding to 278~mm along $ \zeta $, quantities were uniformly extrapolated up to the position of the window, i.e., $ \zeta = \SI{800}{\mm} $, while outside the test chamber, i.e., for $ \zeta > \SI{800}{\mm} $, we assume laboratory conditions with  $ T = \SI{300}{\kelvin} $ and $ p = \SI{1}{atm} $.  
 
In order to reduce the computational effort related to the CFD simulations, the temperature on the surface of the probe was fixed at $ \SI{350}{\kelvin} $. During a PWT experiment, instead, this will rise to an equilibrium value as a result of the heat transfer between the gas and the material. As the thermal boundary layer around the probe is very small ($ \sim \SI{5}{\mm} $) with respect to the jet core ($ \sim \SI{150}{\mm} $), where much of the emission is originating, we expect a limited effect on $ \spectralradiance[g,e] $.

While a detailed validation of the CFD simulations of the plasma flow field is currently missing, the following analysis aims to show the principal effects of plasma emission on the measured surface temperature, providing better understanding and interpretation of the experimental results.

\subsection{Line-of-sight and surface radiance}
\label{sec:TCP_plasma_radiance}
We used the Non-EQuilibrium Air Radiation (NEQAIR v.15.0) code \cite{Cruden2014, NEQAIRweb} to simulate the spectral radiance emitted by the plasma along the line of sight. This allows to compute line-by-line radiation spectra, including spontaneous emission, absorption and stimulated emission, due to transitions between different energy states of atomic and molecular species through a non-uniform gas mixture. Inputs to the code are the gas temperature and number density of constituent species along the optical path, which were provided by the aforementioned CFD simulations. Considering LTE, the population of the excited states is assumed to follow a Boltzmann distribution.

\begin{figure*}[h!]
	\centering
	\includegraphics[trim={0.5cm, 0cm, 0.5cm, 0cm}, clip, width=.98\textwidth]{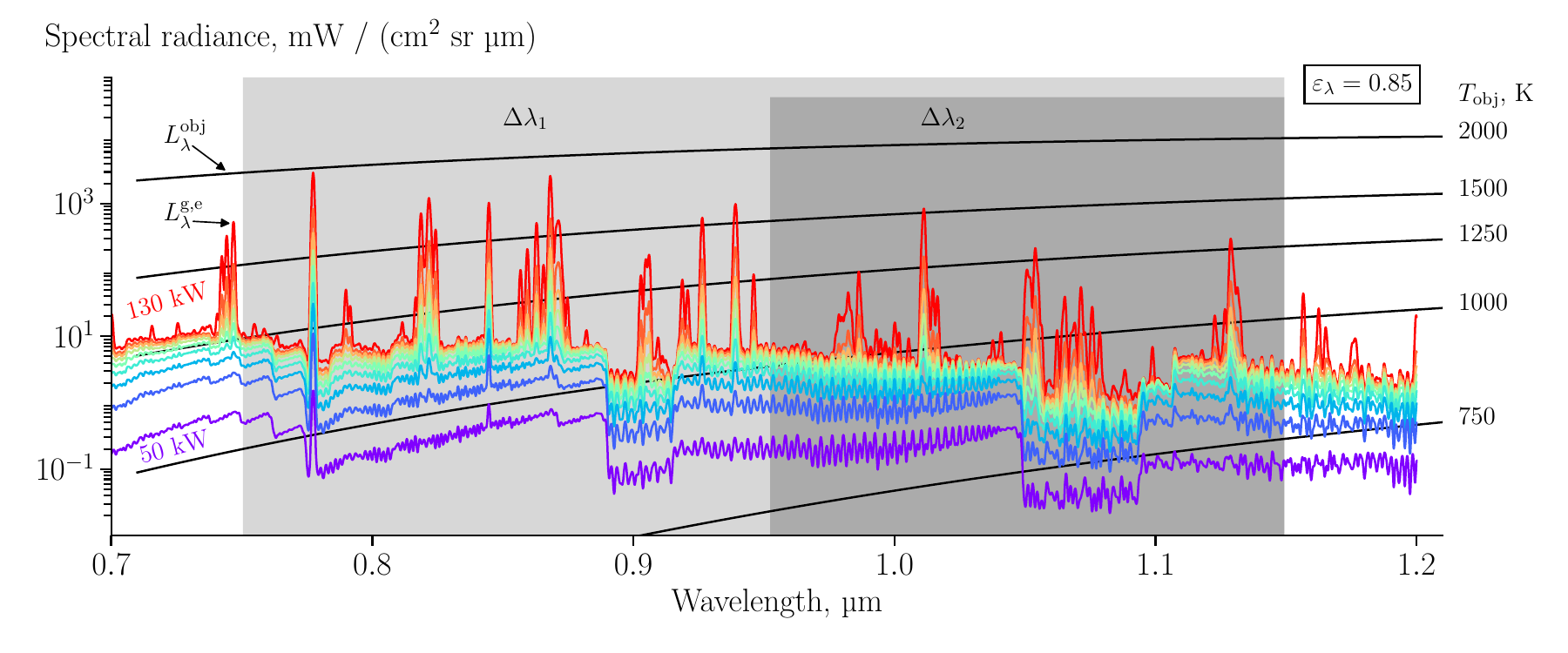}
	\caption[]{Comparison between the simulated values of $ \spectralradiance[g,e] $ and $ \spectralradiance[obj] = \spectraleps \spectralradiance[bb](\Tobj)$, for different values of $ \Tobj $ and $ \varepsilon_{\lambda} = 0.85 $. The emission originating from the plasma along the instrument optical path can significantly interfere with the object's emission up to 1500 K.}
	\label{fig:Lobj_Latm_50mbar_air_Pel50-130kW_Tw750-2000}
\end{figure*}

For different values of $ \Pelsim $, Fig.~\ref{fig:Lobj_Latm_50mbar_air_Pel50-130kW_Tw750-2000} shows the simulated $ \spectralradiance[g,e] $ spectra in the $ \IRMband{0.7}{1.2} $ range. For visualization purposes, the spectral resolution was downgraded through a convolution with a Gaussian lineshape of $ \SI{0.75}{\nm} $ at full-width at half maximum. The lines originating from excited states of $ \ce{O} $ and $ \ce{N} $ atoms are clearly evident, together with the background radiation, mainly due to the first-positive rovibronic transition of $ \ce{N2} $. 
In the same plot, the spectral radiance $ \spectralradiance[obj] = \spectraleps \spectralradiance[bb](\Tobj) $, from a gray-body with $ \spectraleps = 0.85 $, is also shown for different values of the object temperature between $ \SI{750}{\kelvin} $ and $ \SI{2000}{\kelvin} $.

Lastly, the shaded gray areas represent the sensitive wavelength bands of the MR1SB two-color pyrometer. 
We can observe how emission from the plasma easily overcomes the object's radiance up to $ \Tobj = \SI{1000}{\kelvin} $, while the peaks corresponding to the atomic lines achieve comparable values even above $ \Tobj = \SI{1500}{\kelvin} $. 
Additionally, since $ \spectralradiance[g,e] $ is highly non-uniform within the sensitive bands of the two-color pyrometer, the ratio of the detected signals can be strongly affected.

\subsection{Sample visibility factors}

For a range of temperatures  $ \SI{500}{\kelvin} < \Tobj < \SI{2000}{\kelvin} $ and $ \varepsilon\spectral = 0.85 $, we compute the sample band visibility factors, $ \nu\inband[i] $, adapting the definition of Sakuta and Boulos~\cite{Sakuta1988} to consider also the instrument spectral response in each band
\begin{equation}
	\nu\inband[i] = \frac{\int\inband[i] \Rnorm[,i]  \spectralemittance \spectralradiance[bb](\Tobj)} {\int\inband[i] \Rnorm[,i] \left[ \spectralemittance \spectralradiance[bb](\Tobj)  + \spectralradiance[g,e] \right]}.
\end{equation}
These coefficients relate the detected object emission to the sum of the object and plasma emission along the line of sight, clearly showing the plasma interference effect on radiation thermometry.

\begin{figure*}[h!]
	\centering
	\subfigure[]
	{\includegraphics[trim = {0cm, 0.0cm, 0cm, 0cm},clip,width=.45\textwidth]{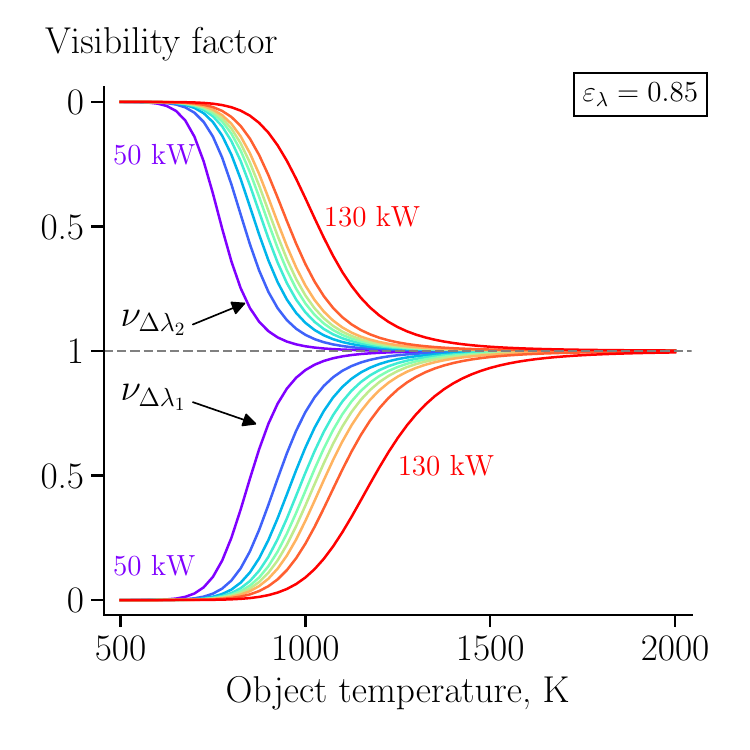}
		\label{fig:TCP_nuw_nun_eps0.8}}
	\subfigure[]
	{\includegraphics[trim = {0cm, 0.0cm, 0cm, 0cm},clip,width=.45\textwidth]{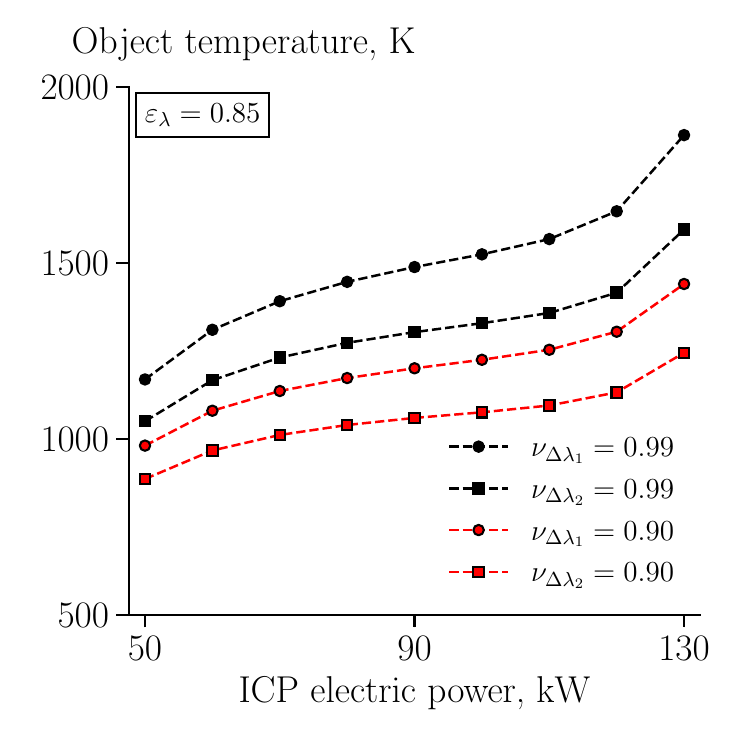}
		\label{fig:TCP_nu99_nu90_Tobj}}
	\caption{(a) Sample visibility factors, $ \nu\inband[1] $ and $ \nu\inband[2] $, as a function of the object temperature, for different values of $ \Pelsim $ and for $ \varepsilon\spectral = 0.85 $. (c) Object temperature for which $  \nu\inband[i] = 0.99 $ and $ \nu\inband[i] = 0.90 $, as a function of $ \Pelsim $, for $ \varepsilon\spectral = 0.85 $. 	}
\end{figure*}

In Fig.~\ref{fig:TCP_nuw_nun_eps0.8} we can appreciate the trend of $ \nu\inband[1] $ and $ \nu\inband[2] $ with $ \Tobj $ for different values of $ \Pelsim $. Both visibility factors are close to zero until $ \SI{750}{\kelvin} $, below which the plasma emission overcomes significantly the object emission and prevents its detection. A transition region is observed for $ \SI{750}{\kelvin} < \Tobj < \SI{1500}{\kelvin} $, where object and plasma emission have similar intensities. Only above $ \SI{1500}{\kelvin} $ the plasma emission becomes negligible and the visibility factors approach one. 
Additionally, Fig.~\ref{fig:TCP_nu99_nu90_Tobj} shows the value of $ \Tobj $ as a function of $ \Pelsim $ for which $ \nu\inband[i] = 0.99 $ and $ \nu\inband[i] = 0.90 $, respectively.
This highlights that, for a certain value of the object temperature, the visibility factors have different values in each band, which can further increase the biasing effect on the measured signal ratio.

\newpage
\subsection{Simulated apparent temperature and measurement error}
Using the instrument response model formulated in sec.~\ref{sec:TCP_IRM}, we simulate the effect of the plasma emission on the measured apparent temperature. For $ \spectraleps~=~0.85 $ and $ \SI{500}{\kelvin}~<~\Tobj~<~\SI{2500}{\kelvin} $, $ \spectralradiance[g,e] $ is inserted in eq.~\ref{eq:TCP_rho_meas} to simulate the measured signal ratio. Then, the apparent temperature is computed from eq.~\ref{eq:TCP_Tapp}. Depending on the value of $ \Pelsim $, Fig.~\ref{fig:Tobj_vs_Tapp_vs_Pel} clearly demonstrates that $ \Tapp $ deviates considerably from $ \Tobj $ when the latter is below 1500~K. Correspondingly, the relative error is computed according to eq.~\ref{eq:TCP_error} and shown in Fig.~\ref{fig:Tobj_vs_err_vs_Pel}. A large systematic error is found for low values of $ \Tobj $, while the steepness of the curves demonstrates the high sensitivity to $ \spectralradiance[g,e] $.

Figure~\ref{fig:Tobj_vs_eps_vs_Pel_NEQAIR_e1} synthetically depicts a map of the error induced by $ \spectralradiance[g,e] $, where each curve represents $ e = 1\% $ as a function of $ \Tobj $ and  $ \varepsilon_{\lambda} $. 
A large portion of the instrument range can be affected by errors larger than 1\%, this being more pronounced for high values of $ \Pelsim $ and low values of $ \spectraleps $. However, we should notice that an instrument with a higher temperature range is typically selected for high values of $ \Pelsim $ to accommodate the higher values also expected for $ \Tobj $. 
Moreover, a horizontal asymptote is reached for $ \spectralemittance \rightarrow 0 $, since the instrument would only detect the plasma radiation in this limit condition. 
This graph is useful to provide a first estimation of the error induced by $ \spectralradiance[g,e] $ on the measured temperature, knowing approximately the expected values of $ \Tobj $ and $ \spectralemittance $ for a selected power condition during the experiment.

\subsection{Simulated effect during transient heating}
\begin{figure*}
	\centering
	\subfigure[]
	{\includegraphics[width=.45\textwidth]{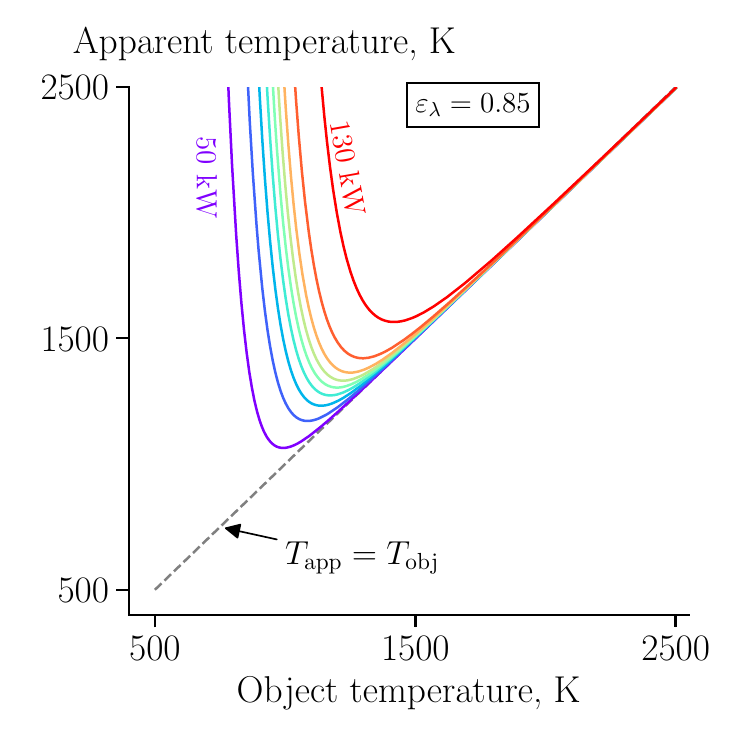} 	\label{fig:Tobj_vs_Tapp_vs_Pel}} 
	\subfigure[]
	{\includegraphics[width=.45\textwidth]{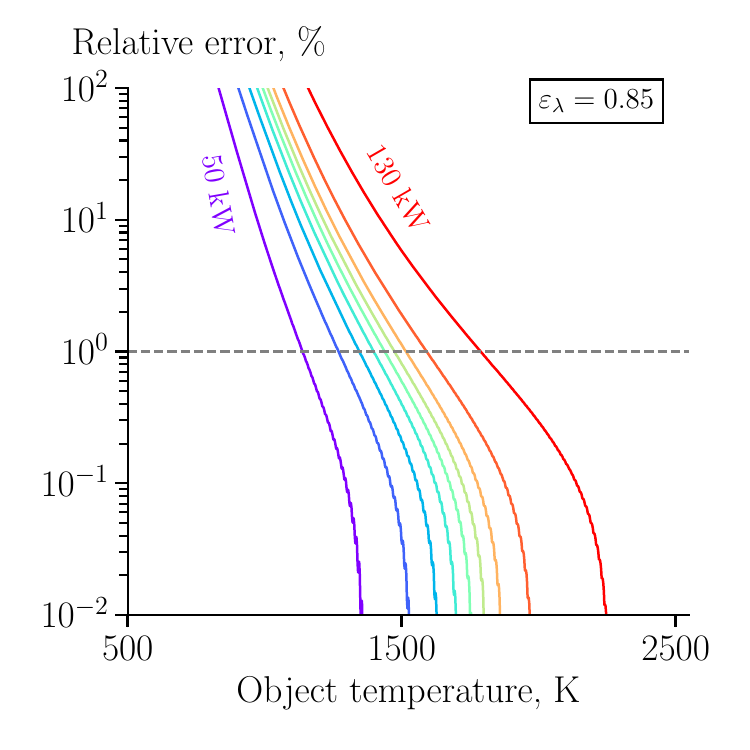} 	\label{fig:Tobj_vs_err_vs_Pel}} \\
	\subfigure[]
	{\includegraphics[width=.48\textwidth]{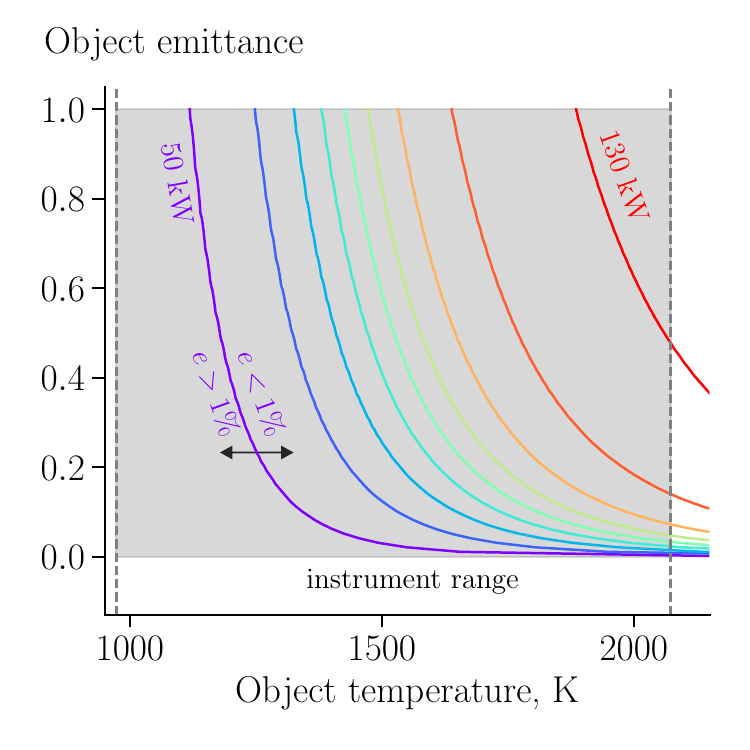}
		\label{fig:Tobj_vs_eps_vs_Pel_NEQAIR_e1} }
		\subfigure[]
	{\includegraphics[width=.48\textwidth]{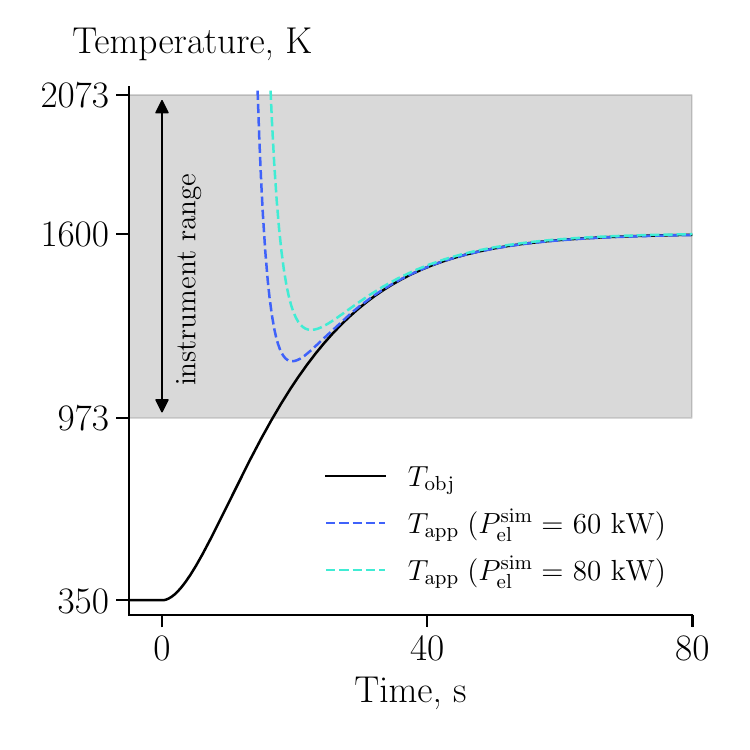} 	\label{fig:Tappsim_transient}} 
	\caption[]{(a) Simulated effect of $ \spectralradiance[g,e] $ on $ \Tapp $ for $ \varepsilon_{\lambda} = 0.85 $, as a function of $ \Pelsim $. (b) Relative error percent of $ \Tapp $ with respect to $ \Tobj $ for $ \varepsilon_{\lambda} = 0.85 $,  as a function of $ \Pelsim $. (c) Map of the simulated measurement error induced by $ \spectralradiance[g,e] $ as a function of $ \Tobj $ and $ \varepsilon_{\lambda} $. Each line represents $ e = 1\% $ and separates the regions for which $ e > 1\% $ and $ e< 1\% $, respectively. (d) Expected material temperature during a PWT experiment, $ T\subrm{obj} $, and correspondent simulated apparent temperature, $ T\subrm{app} $, as it would be measured by the two-color pyrometer considering the influence of $ \spectralradiance[g,e] $ for $ \Pelsim = \SI{60}{\kilo \watt} $ and $ \Pelsim = \SI{80}{\kilo \watt} $. A positive bias induces $ T\subrm{app} $ to decrease during the transient heating and to approach $ T\subrm{obj} $ only when its value is high enough.}
\end{figure*}

Finally, we analyze the effect occurring during the transient heating phase of a material sample for a PWT experiment. As the latter is injected into the plasma flow, we consider the time evolution of the surface temperature to follow the analytical function
\begin{equation}
	T\subrm{obj} (t) =
	\begin{cases}
		T_0 \;\; \textrm{for} \; $ t < \SI{0}{\second} $\\
		T_0 + g \left[1 - (1 + \omega_0 t)\exp(-\omega_0 t)    \right] \;\; \textrm{for} \; $ t > \SI{0}{\second}$,
	\end{cases}
\end{equation}
which represents the step response of a critically dumped second order system, with gain $ g $ and natural frequency $ \omega_0 $. We assume an initial temperature of $ T_0 = \SI{350}{\kelvin} $, while $ g = \SI{1250}{\kelvin} $ and $ \omega_0 = \SI{0.1}{\per \second} $ provide a steady-state temperature of $ \SI{1600}{\kelvin} $ after $ \SI{80}{\second} $. These values are selected to represent qualitatively the experimental results, discussed later in sec.~\ref{sec:TCP_results}. Figure~\ref{fig:Tappsim_transient} shows $ T\subrm{obj}(t) $, along with the simulated behavior of the apparent temperature, $ T\subrm{app}(t) $, as it would be measured by the two-color pyrometer including the effect of $ \spectralradiance[g,e] $ for two representative conditions with $ \Pelsim = \SI{60}{\kilo \watt} $ and $ \Pelsim = \SI{80}{\kilo \watt} $, considering $ \varepsilon_{\lambda} = 0.85 $. The bias induced by the gas radiation along the instrument line of sight causes $ T\subrm{app} $ to drop from large values, rather than to follow the rising trend of $ T\subrm{obj} $, thus resulting in a significant error during the transient heating. As time evolves, the apparent temperature approaches the object temperature when this becomes high enough for $ \spectralradiance[g,e] $ to become negligible.

\section{Experimental results and discussion}
\graphicspath{{figures/}}
\label{sec:TCP_results}

\subsection{Material sample and test conditions}
\begin{figure*}[h!]
	\centering
	\subfigure[]
	{\includegraphics[trim={0cm, 4.5cm, 21.5cm, 0cm}, clip, width=0.29\textwidth]{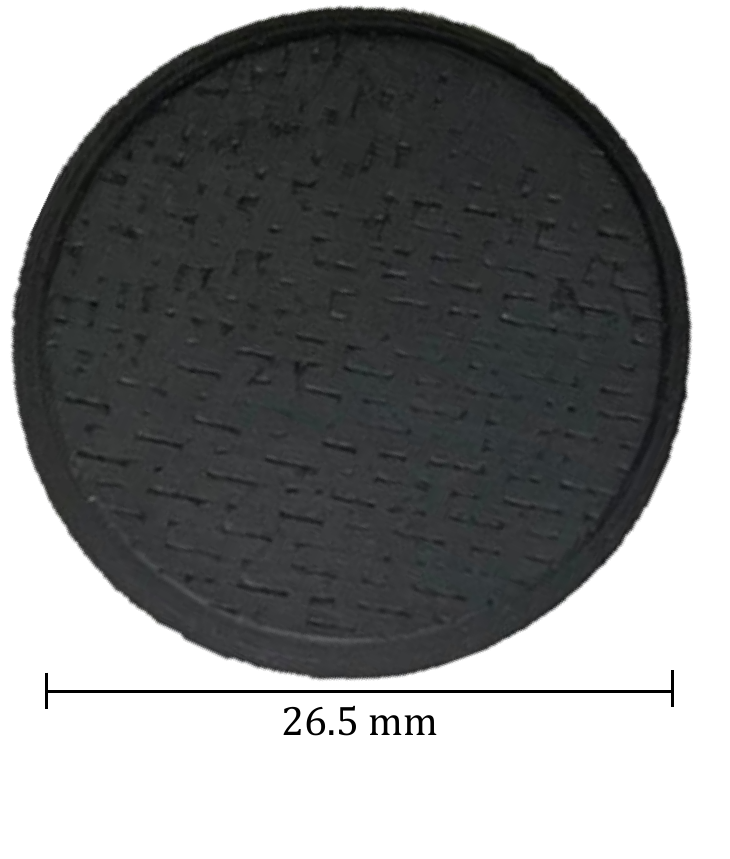}
	\label{fig:MTA_CSiC_CUP_A_pre-test}}
	\subfigure[]
	{\includegraphics[width=0.31\textwidth]{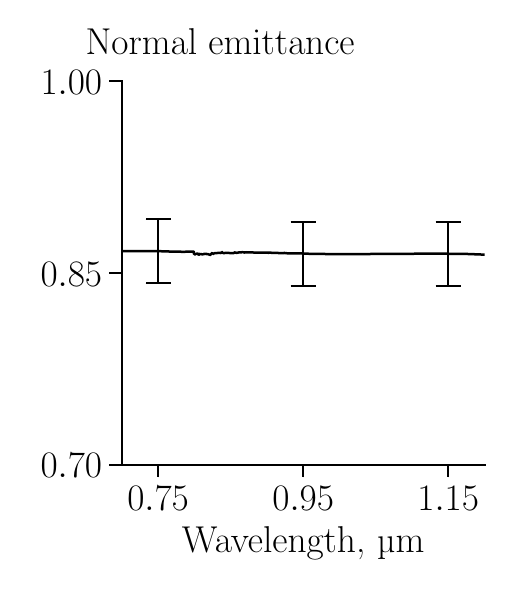}
	\label{fig:MTA_CSiC_CUP_B_eps}}
	\subfigure[]
	{\includegraphics[trim={0cm, 5cm, 21cm, 0cm}, clip, width=0.31\textwidth]{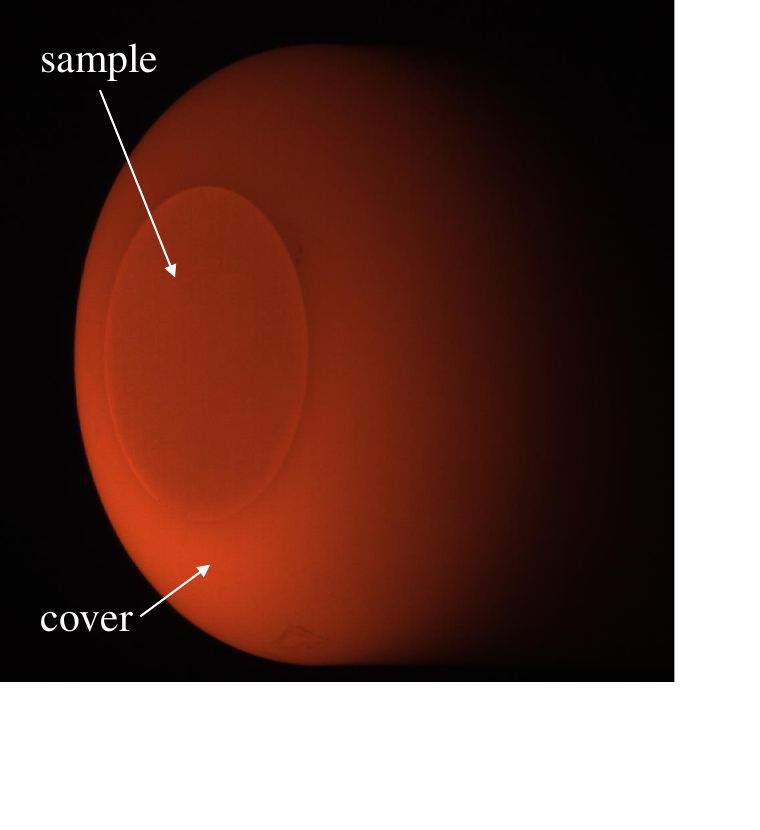}
		\label{fig:MTA_CSiC_CUP_A_test}}
	\caption[]{(a) Picture of the C/SiC sample before the experiment. (b) Normal spectral emittance obtained from room temperature reflectometry of the virgin sample shows $ \spectraleps \approxeq0.85 $, with a uniform behavior within $ \IRMband{0.75}{1.15} $. (c) Picture taken during the plasma exposure, showing the material sample inserted into the probe cover.}
\end{figure*}

Figure~\ref{fig:MTA_CSiC_CUP_A_pre-test} shows the ceramic matrix composite sample (Keraman $ \ce{C/SiC} $, MT-Aerospace) that was selected to study the effect of $ \spectralradiance[g,e] $ on TCP during a PWT experiment. The material is a $ \SI{26.5}{\mm} $ diameter, $ \SI{3}{\mm} $ thick disc, made up of a carbon fiber matrix, with $ \SI{7}{\um} $ diameter filaments, and treated with a polymer vapor infiltration before a final coating ($ \IRMband{60}{80} $ thick) with $ \ce{SiC} $ by chemical vapor deposition. Room temperature spectral reflectance, $ r\spectral $, was characterized before the experiment using a Cary-500 spectrophotometer (Agilent Technologies Inc., USA). The normal spectral emittance in Fig.~\ref{fig:MTA_CSiC_CUP_B_eps}, obtained as $ \spectraleps = 1- r\spectral $, shows a uniform value within $ \IRMband{0.75}{1.15} $, close to 0.85. Hence, this allowed to rely on the gray-body assumption for TCP within the instrument range, as well as to consider a negligible influence of $ \spectralradiance[g,s] $ due to the large value of $ \spectraleps $.
Figure~\ref{fig:MTA_CSiC_CUP_A_test} shows a picture during the plasma exposure, where the material sample was mounted on the $ \SI{50}{\mm} $ diameter ESA standard probe and positioned at $ \SI{385}{\mm} $ from the ICP torch exit.  The experimental values, defining the test conditions achieved in the VKI Plasmatron facility, are reported in table~\ref{tab:test_conditions}. 

\begin{table*}[h!]
	\centering
	\caption{Plasmatron experimental conditions, listing the distance to the torch exit, duration of the plasma exposure, electric power $ \Pel $, test gas mass flow rate $ \mdot $ and chamber pressure $ \pc $.}
		\begin{tabular}{llllll}
			\hline
			distance & duration & gas & $ \Pel $ &  $ \mdot $ & $ \pc $    \\
			 mm       & s             &     & $ \SI{}{\kilo\watt} $ & g/s   & mbar   \\ \hline

			 $ 385 \pm 1 $      & 300   & air & $ 160\pm 10 $ & $ 16 \pm 0.15 $  & $ 50 \pm 1 $  \\
			
			\hline
		\end{tabular}
	\label{tab:test_conditions}   
\end{table*}

\subsection{Two-color pyrometry measurement and correction}

\begin{figure*}[h!]
	\centering
	\subfigure[]
	{\includegraphics[width=0.7\textwidth]{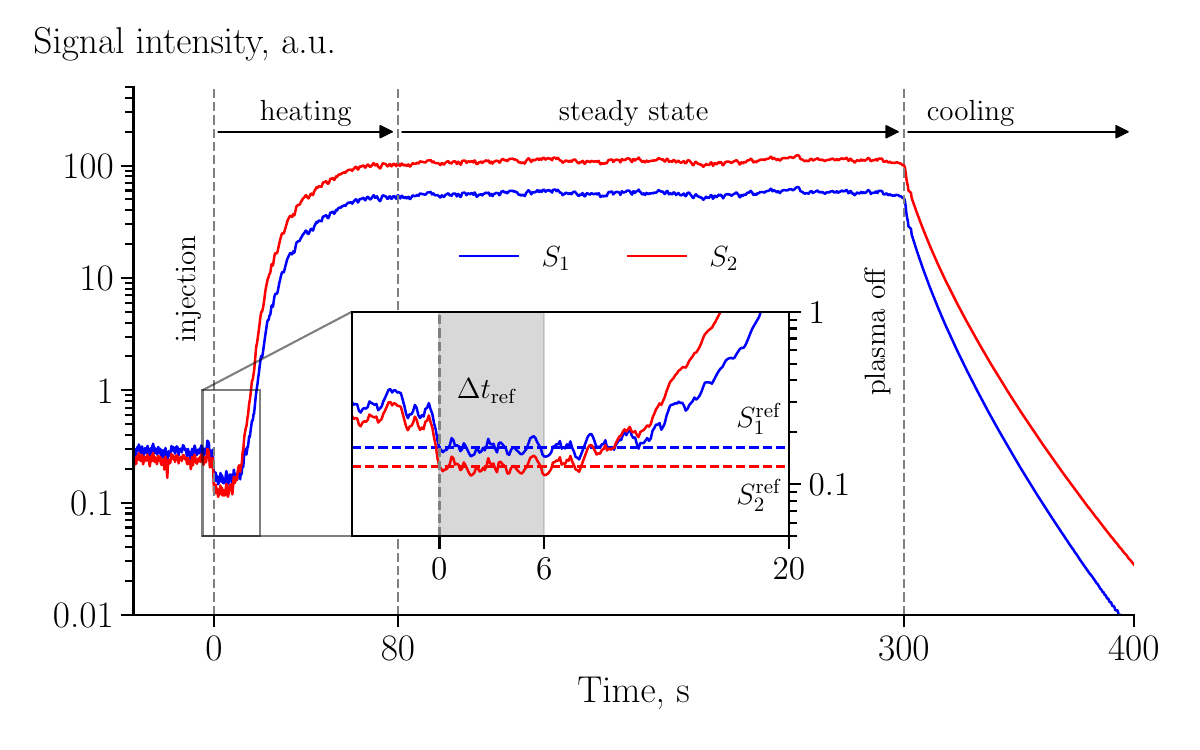}
		\label{fig:MTA_CSiC-CUP-A-T2_signal_intensity}}\\
	\subfigure[]
	{\includegraphics[width=0.60\textwidth]{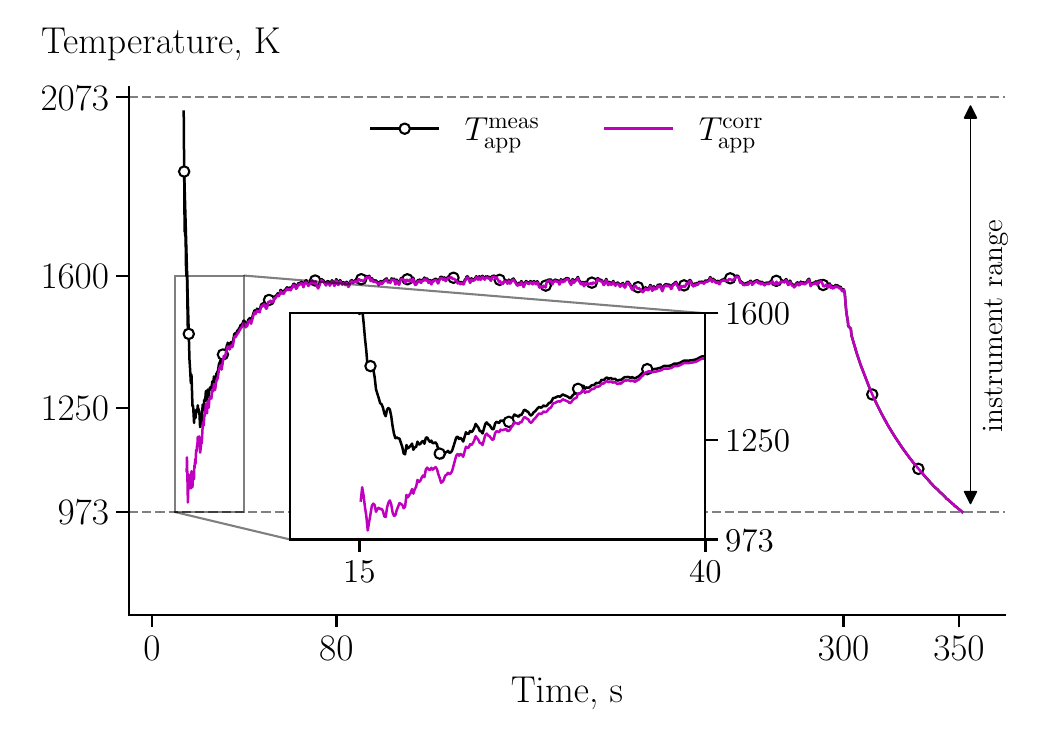}
		\label{fig:MTA_CSiC-CUP-A-T2_Tapp_meas_corr}}
	\subfigure[]
	{\includegraphics[trim = {0cm, 0.cm, 18.5cm, 0cm}, clip, width=0.37\textwidth]{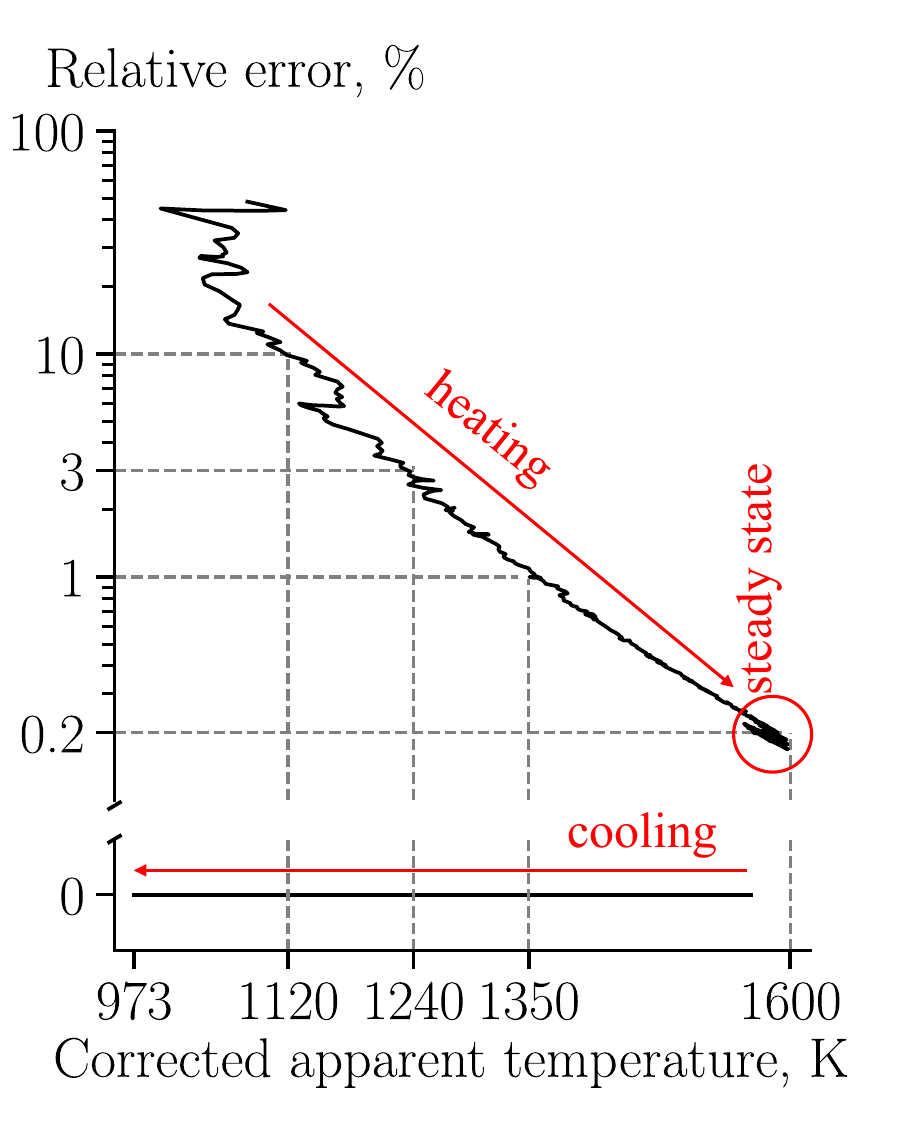}
		\label{fig:MTA_CSiC-CUP-A-T2_err_Tappmeas_Tappcorr}}
	\caption[]{(a) Measured signals $ S_1 $ and $ S_2 $, showing the heating, steady-state and cooling phases. 
		The reference signals for correction, $ S_{1}^{\textrm{ref}} $ and $ S_{2}^{\textrm{ref}} $, are obtained by averaging the values within $ \SI{0}{\second} < t < \SI{6}{\second} $. (b) Comparison between the measured apparent temperature and its corrected value. Notice how $ T\subsuprm{app}{meas} $ decreases during the heating phase, similarly to what has been shown in our simulations (Fig.~\ref{fig:Tappsim_transient}). $ T\subsuprm{app}{corr} $, instead, recovers an increasing trend, as expected.   (c) Relative error between the measured and corrected apparent temperature. $ T\subsuprm{app}{meas} $ falls within $ 1\% $ of $ T\subsuprm{app}{corr} $ only above $ \SI{1350}{\kelvin} $, while their difference is negligible at steady state. Since no correction is applied during the cooling phase, their values coincide.}
\end{figure*}

The time evolution of the measured signals from the two wavelength bands, $ S_1 $ and $ S_2 $, is shown in Fig.~\ref{fig:MTA_CSiC-CUP-A-T2_signal_intensity}.
Here, $ t = \SI{0}{\second} $ represents the instant when the sample is injected into the plasma flow. After a transient heating phase of approximately $ \SI{80}{\second} $, the signals settle to a steady value; then, the plasma torch is switched off at $ t = \SI{300}{\second} $, allowing to record the sample cool down until $ t = \SI{400}{\second} $. 

We can notice that both $ S_1 $ and $ S_2 $ drop at $ t=\SI{0}{\second} $, as the sample  obstructs the lower half of the plasma jet after injection.
Moreover, shortly after $ t = \SI{0}{\second} $ we can consider $ \Tobj $ to be low enough such that $ \spectralradiance[g,e] \gg \spectralradiance[obj] $ and  $ \spectralradiance \approxeq \spectralradiance[g,e] $.
We indicate with $ S\subsuprm{1}{ref} $ and $ S\subsuprm{2}{ref} $ the time-averaged signals within a reference interval $ \Delta t_{\textrm{ref}} $ shortly after the injection time, in this case between $ \SI{0}{\second} $ and $ \SI{6}{\second} $. This range should be chosen according to the particular measurement condition and must end before the signals start to rise due to the emission from the object's surface. Since $ \spectralradiance \approxeq \spectralradiance[g,e] $ within $ \Delta t_{\textrm{ref}} $, then eq.~\ref{eq:TCP_signal} yields
\begin{equation}
 	S_{i}^{\textrm{ref}} \approxeq k_i \throughput \int\inband[i] \Rnorm[,i] \spectralradiance[g,e] d\lambda.
 \end{equation}
Assuming that $ \spectralradiance[g,e] $ does not vary significantly during the sample heating and steady-state phases, the corrected signals
 \begin{equation}
 	S_{i}^{\textrm{corr}} =
 	\begin{cases}
 		S_i - S_{i}^{\textrm{ref}} \approxeq k_i \throughput \int\inband[i] \Rnorm[,i] \spectraleps \spectralradiance[bb] (\Tobj) d\lambda \;\; \textrm{for} \; 0\leq t \leq \SI{300}{\second} \\
 		S_i \;\; \textrm{for} \; t > \SI{300}{\second}
 	\end{cases}
 	\label{eq:S_corr}
 \end{equation}
approximate the contributions originating from the object's surface only. Notice that a correction is not necessary after 300~s, as $ \spectralradiance[g,e] = 0 $ after the plasma is switched off.

The measured apparent temperature, $ T\subsuprm{app}{meas} $, is obtained considering the ratio $ \rho\subrm{m}~=~S_1/S_2 $ and applying the calibration curve obtained in sec.~\ref{sec:TCP_calibration} as $ T\subrm{app}~=~\Hfunction\subsuprm{c}{-1}(\rho\subrm{m}) $.
The corrected ratio $ \rho\subsuprm{m}{corr} = S_{1}^{\textrm{corr}} / S_{2}^{\textrm{corr}} $, instead, can be used to determine a corrected value of the apparent temperature $ T\subsuprm{app}{corr} = \Hfunction\subsuprm{c}{-1}(\rho\subsuprm{m}{corr}) $. The latter represents the best estimate of the object temperature, removing the biasing effect of the plasma emission along the line-of-sight.
Figure~\ref{fig:MTA_CSiC-CUP-A-T2_Tapp_meas_corr} shows the measured and corrected temperature values. 
We can notice how, before correction, $ T\subsuprm{app}{meas} $ follows a similar trend to those predicted by our simulations in Fig~\ref{fig:Tappsim_transient}. Namely, the value drops from the top of the instrument range, to rise again only after $ t \approxeq  \SI{20}{\second} $. Hence, this peculiar trend can be associated to the biasing effect of $ \spectralradiance[g,e] $ on the measured signal.
The corrected temperature, instead, recovers a rising trend, as it is expected during the transient heating of the material. At steady state, the surface temperature is high enough for $ \spectralradiance[g,e] $ to have a small effect and the correction has a negligible impact. 

Considering $ T\subsuprm{app}{corr} $ to be best representative of the material surface temperature, then $ e = (T\subsuprm{app}{meas} - T\subsuprm{app}{corr}) / T\subsuprm{app}{corr} $ represents the relative error of the measured value with respect to the corrected one. Figure~\ref{fig:MTA_CSiC-CUP-A-T2_err_Tappmeas_Tappcorr} shows that  $ T\subsuprm{app}{meas} $ falls within $ 1\% $ of $ T\subsuprm{app}{corr} $ only above $ \SI{1350}{\kelvin} $, while the error is larger than $ 10\% $ below $ \SI{1120}{\kelvin} $. Their difference becomes negligible ($ 0.2 \% $) at steady state and, since no correction is applied after the plasma is switched off, their values coincide during the cooling phase.

\subsection{Comparison to spectrometry}

\begin{figure}[h!]
	\centering
	\subfigure[]
	{\includegraphics[width=0.99\linewidth]{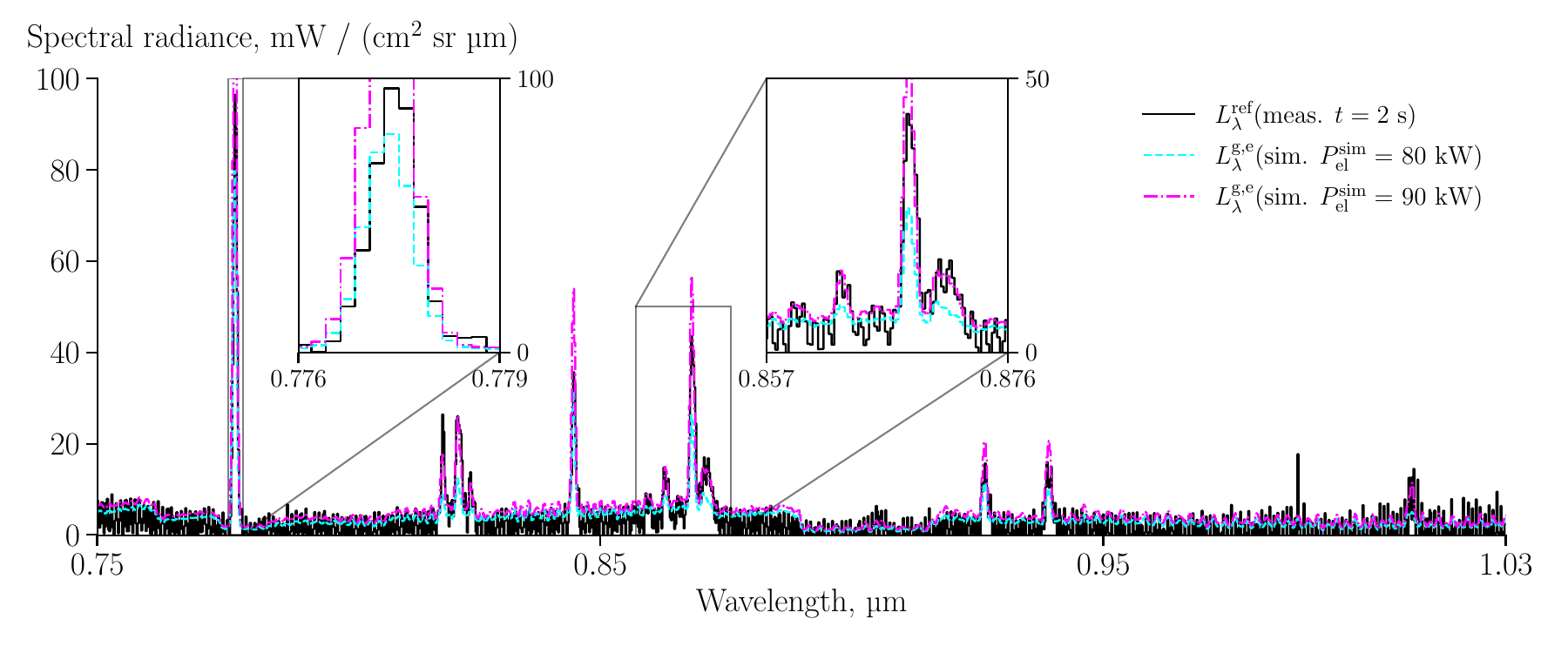} 	\label{fig:MTA_CSiC-CUP-A-T2_L_ge_meas_sim}}
	\subfigure[]
	{\includegraphics[width=0.45\textwidth]{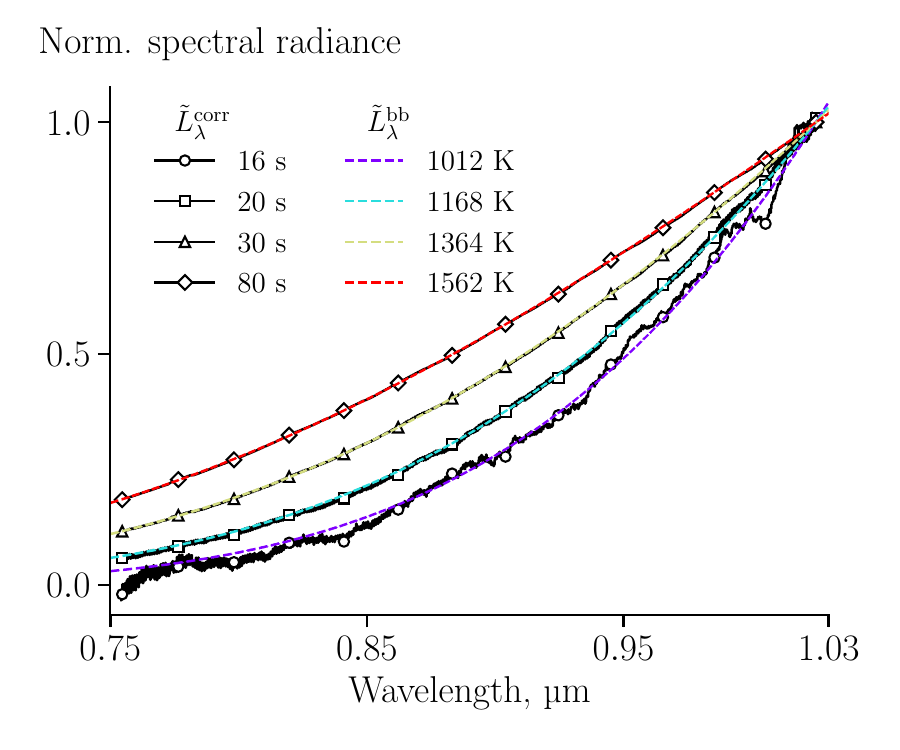}
		\label{fig:MTA_CSiC-CUP-A-T2_L_norm}}
	\subfigure[]
	{\includegraphics[width=0.45\textwidth]{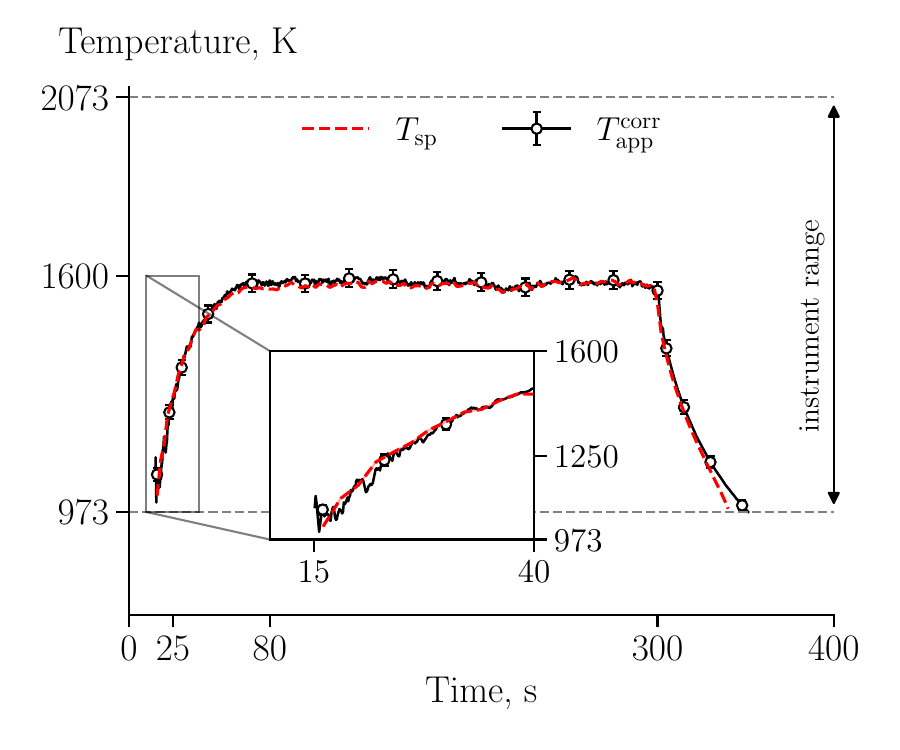}
		\label{fig:MTA_CSiC-CUP-A-T2_Tappcorr_Tsp}}
	\caption{(a) Reference spectrum, $ \spectralradiance[ref] $, measured at $ t = \SI{2}{\second} $, representing the spectral radiance of the gas along the optical path. Simulated $ \spectralradiance[g,e] $ spectra for $ \Pelsim = 80 $ and $ \SI{90}{\kW} $, obtained in sec.~\ref{sec:TCP_plasma_radiance}, provide comparable intensities. (b) Normalized corrected spectral radiance of the object surface, $ \tilde{L}\subsuprm{\lambda}{corr} $, for different time instants after injection. These are fitted with the normalized Planck distribution, $ \tilde{L}\subsuprm{\lambda}{bb}(T) $, to provide the temperature $ T\subrm{sp} $. (c) Comparison between  $ T\subrm{sp} $ and the corrected apparent temperature from the two-color pyrometer, $ T\subsuprm{app}{corr} $, show noticeable agreement.}
\end{figure}

The spectrometer, described in sec.~\ref{sec:TCP_setup_spectro}, allowed to probe the spectral radiance emitted by the object's surface and gas along the line of sight. 
Following the previous discussion, since we expect $ \spectralradiance[obj] \ll \spectralradiance[g,e] $ for $ \SI{0}{\second} < t < \SI{6}{\second} $, a reference spectrum $ \spectralradiance[ref] $ within this time interval is such that $ \spectralradiance[ref] \approxeq \spectralradiance[g,e] $.
Figure~\ref{fig:MTA_CSiC-CUP-A-T2_L_ge_meas_sim} compares $ \spectralradiance[ref] $ at $ t = \SI{2}{\second} $ to the values of $ \spectralradiance[g,e] $, computed in sec.~\ref{sec:TCP_plasma_radiance} for $ \Pelsim = \SI{80}{\kW} $ and $ \Pelsim = \SI{90}{\kW} $, demonstrating that our radiative transfer simulations provide comparable intensities to those observed during actual experiments. 

To measure the surface temperature from the observed spectra, these are first corrected for the reference spectrum as 
	\begin{equation}
		\begin{cases}
			\spectralradiance[corr] = \spectralradiance[m] - \spectralradiance[ref] \approxeq \spectraleps \spectralradiance[bb] (\Tobj) \;\; \textrm{for} \; 0\leq t \leq \SI{300}{\second} \\
			\spectralradiance[corr] = \spectralradiance[m] \;\; \textrm{for} \; t > \SI{300}{\second}.
		\end{cases}
		\label{eq:L_corr}
	\end{equation}
Then, employing a similar method to the one shown by Savino~et~al.~\cite{Savino2020}, the material's surface temperature can be obtained by a least-square fitting of the normalized spectrum $ \tilde{L}\subsuprm{\lambda}{corr} = \spectralradiance[corr] / L_{\lambda_{\textrm{norm}}}^{\textrm{corr}} $ with the normalized Planck distribution $ \tilde{L}\subsuprm{\lambda}{bb}(T)~=~\spectralradiance[bb](T)/L_{\lambda_{\textrm{norm}}}^{\textrm{bb}}(T)  $. For this procedure we consider $ \lambda\subrm{norm} = \SI{1.03}{\um} $, while the fitting provides the temperature $ T\subrm{sp} $. 
Figure~\ref{fig:MTA_CSiC-CUP-A-T2_L_norm} shows some representative $ \tilde{L}\subsuprm{\lambda}{corr} $ spectra at different times, together with the fitted $ \tilde{L}\subsuprm{\lambda}{bb}(T\subrm{sp}) $ spectra, and the corresponding temperatures. Before $ t = \SI{16}{\second} $, however, the signal is too low to provide reliable measurements.
Figure~\ref{fig:MTA_CSiC-CUP-A-T2_Tappcorr_Tsp} compares the measured surface temperature by means of the spectroscopic method, $ T\subrm{sp} $, together with the corrected value from the two-color ratio pyrometry, $ T\subsuprm{app}{corr} $, obtained in the previous section. 
Their values show a close agreement, confirming that the subtraction of the reference signals correspondent to the line-of-sight plasma radiation allows to recover a reliable measurement of the material surface temperature during the transient heating phase.

\section{Conclusions}

Two-color ratio pyrometry provides powerful in situ diagnostics for plasma wind tunnel experiments of aerospace materials. However, the intense plasma radiation along the optical path poses a major concern for measurement accuracy.

In this work, we formulated a model of the instrument response that allowed a numerical assessment of the error. The gas spectral radiance was computed with a radiation code, using temperature, pressure, and gas composition provided by CFD simulations of the plasma flow field. Our results show that gas radiation reaches a comparable intensity to the one emitted by the material for typical surface temperatures expected during PWT experiments. As a result, the measured signal ratio is shifted with respect to the value obtained during calibration, causing large positive errors in the measured temperature. The effect is particularly relevant during the transient heating phase of the material sample exposed to the plasma jet, while steady-state values appear negligibly affected.

A plasma wind tunnel test on a ceramic matrix composite material provided experimental data that confirmed the predicted trends, highlighting that a correction is necessary to improve the accuracy. Subtraction of reference signals, obtained right after the injection time, revealed a valuable procedure for a high-emittance material, as those could be interpreted as the contributions originating from the gas radiation only. A spectrometer, simultaneously pointed at the material, provided the spectral signature of either the gas and the surface, confirming the correction.

With the presented work, we provide a methodology to predict and understand an important biasing effect on ratio pyrometry, extending previous analyses concerning the influence of different environmental factors, and allowing experimentalists to achieve improved measurements in applications involving high-temperature gases. While a measurement correction is possible, future work should focus on identifying different wavelength ranges to minimize the interference of the plasma radiation. In this case, a compromise with the instrument's sensitivity needs to be accounted for. Additionally, while high-emittance materials allowed to neglect reflections from the surrounding gas, these could become important and would require further consideration in the case, for instance, of low-emittance metallic samples.

\section*{Acknowledgments}
The research of A. Fagnani was funded by the Research Foundation - Flanders (dossier n. 1SB3121N).
The experimental activities of this work were supported by the Air Force Office of Scientific Research (Grant N. FA9550-18-1-0209). 
V.Romano is acknowledged for the mesh convergence study on the CF-ICP code simulations.
The authors would like to thank J. Freitas Monteiro for the precious help as Plasmatron test engineer and P. Collin as Plasmatron technical operator.
S. Smolka, B.Bras and G.van Papendrecht are acknowledged for the support with the reflectometry measurements at the ESA ESTEC laboratories.

\section*{Declaration of Competing Interest}
The authors declare that they have no known competing financial interests or personal relationships that could have appeared to influence the work reported in this paper.

\bibliographystyle{elsarticle-num} 

\bibliography{references}

\end{document}